\documentclass[aip,reprint]{revtex4-1}

\usepackage{amsmath}
\usepackage{amsfonts}
\usepackage{refstyle}
\usepackage{chngcntr}
\usepackage{verbatim}
\usepackage{braket}
\usepackage{graphicx}
\usepackage{rotating}
\usepackage[flushleft]{threeparttable}

\usepackage[version=3]{mhchem}

\usepackage{subfig}

\newcommand*{\mr}{\mathrm}
\newcommand*{\bW}{{\bf W}}
\newcommand*{\bG}{{\bf G}}

\newcommand*{\ti}{\tilde}
\newcommand*{\mca}{\mathcal{A}}
\newcommand*{\br}{{\bf r}}
\newcommand*{\bk}{{\bf k}}

\newcommand*\citeref[1]{ref. \citenum{#1}}
\newcommand*\citerefs[1]{refs. \citenum{#1}}
\newref{eq}{name={eq~},names={eqs~},Name={Eq~},Names={Eqs~},refcmd={(\ref{#1})}}

\begin{document}

\author{Elvar \"{O}. J\'{o}nsson}
\affiliation{COMP Centre of Excellence and Department of Applied Physics, Aalto University
  School of Science, P.O. Box 11100, FI-00076 Aalto, Espoo, Finland}

\author{Susi Lehtola}
\affiliation{COMP Centre of Excellence and Department of Applied Physics, Aalto University
  School of Science, P.O. Box 11100, FI-00076 Aalto, Espoo, Finland}
\affiliation{Chemical Sciences Division, Lawrence Berkeley National
  Laboratory, Berkeley, California 94720, United States}

\email{susi.lehtola@alumni.helsinki.fi}

\author{Martti Puska}
\affiliation{COMP Centre of Excellence and Department of Applied Physics, Aalto University
  School of Science, P.O. Box 11100, FI-00076 Aalto, Espoo, Finland}

\author{Hannes J\'{o}nsson}
\affiliation{COMP Centre of Excellence and Department of Applied Physics, Aalto University
  School of Science, P.O. Box 11100, FI-00076 Aalto, Espoo, Finland}
\affiliation{Faculty of Physical Sciences, University of Iceland,
  107 Reykjav\'ik, Iceland}

\title{Theory and applications of generalized Pipek--Mezey Wannier functions}


\begin{abstract}
The theory for the generation of Wannier functions within the
generalized Pipek--Mezey approach [Lehtola, S.; J\'onsson, H. {\em
    J. Chem. Theory Comput.} {\bf 2014}, 10, 642] is presented and an
implementation thereof is described.  Results are shown for systems
with periodicity in one, two and three dimensions as well as isolated
molecules. The generalized Pipek--Mezey Wannier functions (PMWF) are
highly localized orbitals consistent with chemical intuition where a
distinction is maintained between $\sigma$- and $\pi$-orbitals. The
PMWF method is compared with the so-called maximally localized Wannier
functions (MLWF) that are frequently used for the analysis of
condensed matter calculations. Whereas PMWFs maximize the localization
criterion of Pipek and Mezey, MLWFs maximize that of Foster and Boys
and have the disadvantage of mixing $\sigma$- and $\pi$-orbitals in
many cases.  The PMWF orbitals turn out to be as localized as the MLWF
orbitals as evidenced by cross-comparison of the values of the PMWF
and MLWF objective functions for the two types of orbitals.  Our
implementation in the atomic simulation environment (ASE) is
compatible with various representations of the wave function,
including real-space grids, plane waves and linear combinations of
atomic orbitals. The projector augmented wave formalism for the
representation of atomic core electrons is also supported.  Results of
calculations with the GPAW software are described here, but our
implementation can also use output from other electronic structure
software such as ABINIT, NWChem and VASP.
\end{abstract}

\maketitle


\section{Introduction}

Thanks to the continuing development of approximate density
functionals in the past decades, Kohn--Sham density functional
theory\cite{Hohenberg1964, Kohn1965} (KS-DFT) has become an essential
workhorse of present-day computational chemistry and solid-state
physics,\cite{Becke2014} allowing e.g. for the \emph{in silica} design
of new materials\cite{Jain2016}. KS-DFT calculations are based on the
variational minimization of an energy functional
\begin{equation} \label{eq:Efunc}
E = E[n_\alpha, n_\beta]
\end{equation}
that depends on the spin-up and spin-down electron densities
$n_\alpha$ and $n_\beta$. In KS-DFT, the spin-$\sigma$
density is represented by a fictitious system of non-interacting
electrons as
\begin{equation} \label{eq:density}
  n_\sigma ({\bf r}) =  \sum_{i\text{ occ.}} \left| \phi_{i\sigma} ({\bf r}) \right|^2,
\end{equation}
which amounts to writing the wave function of the fictitious system as
an antisymmetrized product i.e. a Slater determinant of
non-interacting single-particle wave functions $\phi_{i \sigma}$ that
are referred to as spin-orbitals. During the minimization of
\eqref{Efunc} with respect to the \eqref{density}, the spin-orbitals
are typically chosen to be the ones that diagonalize the effective
one-particle Hamiltonian operator, because this leads to a decoupling
of the differential equations for the optimal orbitals
as\cite{Kohn1965}
\begin{equation} \label{eq:ks}
 \left[ -\frac 1 2 \nabla^2 + V({\bf r}) + v_\sigma^\text{xc}({\bf r})
   \right] \phi_{i\sigma}({\bf r}) = \epsilon_{i \sigma} \phi_{i\sigma}({\bf
   r}).
\end{equation}
Here, $V({\bf r})$ is the external potential generated by the nuclei
and the classical Coulomb potential generated by \emph{all} the
electrons, whereas $v_\sigma^\text{xc}$ is the exchange-correlation
potential which maps the fictitious non-interacting system onto the
real, interacting system of electrons, which is not known in general
but for which many approximations have been developed over the
years\cite{Becke2014}. The orbitals obtained from \eqref{ks} are
referred to as the {\em canonical orbitals} (COs), or {\em
crystal(line) orbitals} in periodic systems.

As seen from \eqref{density}, each electron in the system occupies
some orbital and is seen to interact only with the average field
generated by the nuclei and all the electrons in the system, since the
energy functional, \eqref{Efunc}, only depends on the \emph{total}
spin densities $\{ n_\sigma \}$.  However, the COs typically extend
over the whole system and \emph{all} the orbitals therefore contribute
to e.g. \emph{any} covalent chemical bond between two atoms.  But, the
occupied orbitals in \eqref{Efunc, density} are arbitrary in the sense
that any unitary transformation thereof, $\{ \sum_{j\text{ occ}}
W^\sigma_{ji} \phi_{j \sigma} \}$, yields the same $n_\sigma$ as $\{
\phi_{i\sigma} \}$ if $({\bf W}^\sigma)^\dagger = ({\bf
  W}^\sigma)^{-1} $. Once the COs have been determined, one can thus
make a rotation of the occupied orbitals through some ${\bf W}^\sigma$
that results in \emph{localized orbitals} (LOs), which yield the same
physical description of the system as the COs (as they span the same
spin densities) but are confined to a smaller region of space. Because
LOs can typically be interpreted in terms of chemical bonds, they can
provide useful information about chemical interactions as well as a
rationalization of the geometrical arrangement of atoms in molecules
and solids\cite{Silvi2011, *Vidossich2014}.

The same discussion as above for KS-DFT also applies to Hartree--Fock
(HF) theory in that the HF orbitals are delocalized but the total
energy is invariant under rotation of the occupied orbitals, and more
meaningful LOs can therefore be formed with post-processing of HF
calculations. Unfortunately, there is no physical principle within
KS-DFT or HF that would guide the choice of the LOs, so several
different schemes have been suggested. The three that are most often
used for isolated systems are the one proposed by Foster and Boys
(FB), who suggested minimizing the second moment (i.e. variance) of
the orbital densities\cite{Foster1960}, the one proposed by Edmiston
and Ruedenberg (ER), who suggested maximizing the self-Coulomb
repulsion of the orbital densities\cite{Edmiston1963}, and the one
proposed by Pipek and Mezey (PM), who suggested maximizing Mulliken
partial charge estimates\cite{Mulliken1955} per orbital on the atoms
of the system\cite{Pipek1989}.  Other schemes that have been suggested
include the von Niessen (vN) scheme\cite{VonNiessen1972}, which
maximizes the density self-overlap, the fourth moment (FM)
scheme\cite{Hoyvik2012a}, which minimizes the fourth moment of the
orbital spread (a variation on the FB scheme), and the natural bond
orbital (NBO) method\cite{Reed1985a, *Weinhold2016}, which relies on
the diagonalization of localized blocks of the one-electron density
matrix.  For the purpose of generating LOs, the various localization
procedures usually produce similar results. However, one important
advantage of the PM and NBO schemes over the others is that a proper
separation of $\sigma$- and $\pi$-bond orbitals is
maintained\cite{Pipek1989, Reed1985a}.  The generation of LOs from HF
orbitals (which are usually similar to KS-DFT orbitals) has been
discussed recently by H{\o}yvik and coworkers\cite{Hoyvik2013,
  *Hoyvik2013a, *Hoyvik2014, *Hoyvik2016}.

While a multitude of localization methods exists for molecular
systems, the available selection for solid state systems is
more limited. In fact, it appears that the condensed matter community
generally associates LOs in extended systems with the ``maximally
localized Wannier function'' approach\cite{Marzari1997, Marzari2012},
which is an adaption of the original FB approach\cite{Foster1960}
to periodic systems\cite{Marzari1997}.  Non-iterative\cite{Lu2004a,
  *Chan2007, *Qian2008} as well as iterative\cite{Yao2010} projection
approaches for obtaining LOs in condensed matter calculations that are
based on quasiatomic orbitals have also been presented in the
literature, but they do not appear to have become widely used.

Here, it must be noted that the name ``maximally localized Wannier
functions'' is rather unfortunate, because it implies that there is a
unique, unambiguous way of determining the locality of
orbitals. However, as discussed above and in \citeref{Marzari2012},
orbital localization is not based on a single, rigorous, unifying
physical principle. All of the localization methods listed above --
the FB, ER, PM, vN, and FM approaches -- rely on the optimization of
some objective function, thus yielding maximally localized orbitals
\textbf{as determined by that objective function}. Instead,
``maximally localized Wannier functions'' should rather be called
Foster--Boys Wannier functions (FBWF), and we will use that convention
throughout the rest of this article.  Furthermore, as it has been
shown recently that more localized \emph{sets} of molecular orbitals
than the ones reproduced by FB can be obtained by optimizing higher
powers of the orbital spreads ($p>1$, whereas FB uses $p=1$), and/or
by using the FM objective function instead of the FB
one,\cite{Hoyvik2012a} it is evident that more localized Wannier
functions than the ones produced by FBWF could be achieved.

A disadvantage of the FBWF approach is the mixing of $\sigma$ and
$\pi$ orbitals. But, any scheme that can reproduce localized orbitals
for molecules can also be used to produce Wannier
functions\cite{Wannier1937}, i.e. LOs for a periodic system, in
contrast to the Bloch states\cite{Bloch1929} which constitute the
extended COs. So, there is no reason why some of the other
localization methods discussed above could not be adapted to periodic
systems.  While the ER, vN, and FM approaches tend to mix $\sigma$ and
$\pi$ orbitals alike to the FB approach, the PM approach avoids such
mixing and is often used for molecular systems. While PM has been used
in $\Gamma$-point calculations of periodic systems based on localized
basis sets and Mulliken analysis\cite{Berghold2000}, a general
formulation of the procedure for FD or PW based calculations has not
been presented. Such a formulation can be problematic since Mulliken
charges (and the related L\"owdin charges) do not have a complete
basis set limit.\cite{Lehtola2014}.

Modifications of the original PM scheme have, however, recently been
presented by Lehtola and J\'onsson in \citeref{Lehtola2014}, where the
Mulliken partial charge estimate can be replaced by a variety of
different, mathematically well-founded alternatives based on e.g. a
real-space division of the system into atomic regions. The schemes
studied in \citeref{Lehtola2014} included the Voronoi
(a.k.a. Wigner--Seitz in periodic systems) partitioning,
Hirshfeld\cite{Hirshfeld1977} or iterative Stockholder
partitioning\cite{Lillestolen2008, Lillestolen2009}, and
Bader\cite{Bader1990} partitioning, among others. It was shown in
\citeref{Lehtola2014} that similar mathematically better-defined
modifications of the PM objective function had been proposed earlier
in the literature.  Cioslowski used a localization scheme based on
Bader charges,\cite{Cioslowski1991} and Alcoba et al. later replaced
them with ``fuzzy atom'' charges \cite{Alcoba2006}. Neither scheme has
become widely used, which we attribute to the lack of the connection
to the PM method that was established in
\citeref{Lehtola2014}. Simultaneously to our work in
\citeref{Lehtola2014}, another modification of the Pipek--Mezey method
using charge estimates based on free-atom orbitals (``intrinsic atomic
orbitals'', IAOs) was proposed by Knizia\cite{Knizia2013}, who coined
the resulting LOs as ``intrinsic bond orbitals'' (IBOs).

An important aspect of the results presented in \citeref{Lehtola2014}
is the realization that the orbitals obtained from a PM localization
are remarkably insensitive to the choice of the partial charge
estimate.  Even qualitatively different estimates were found to yield
similar LOs. Furthermore, for all choices of the partial charge
estimate, a clear separation of $\sigma$- and $\pi$-bonds is
maintained \cite{Lehtola2014}. Since similar LOs can be obtained with
a variety of dissimilar partial charge estimates, the original
Pipek--Mezey method based on Mulliken charges as well as the schemes
by Cioslowski, Alcoba et al. and Knizia can all be considered to be
examples of a generalized PM method described in
\citeref{Lehtola2014}.

Due to the robustness of the generalized PM method, there is
significant flexibility in the choice of the partial charge estimate and
this can be used to ensure that the localization procedure is mathematically rigorous
(thus guaranteeing well-defined basis set limits and robust convergence
of the optimization) and also computationally convenient. By choosing the
partial charges in the generalized PM scheme to be defined as
integrals over the electron density in real-space, the approach can be
ideally suited for condensed phase calculations.

Condensed matter simulations are typically based on a plane wave (PW)
description of the electronic structure, while uniform real-space
grids employing the finite difference (FD) approximation are becoming
increasingly popular. Unlike the original procedure based on Mulliken
charges that does not allow for straightforward application to PW or
FD based calculations, the real-space choice for the partial charges
does not depend on the use of an atom-localized basis set and thus can
be readily implemented in PW or FD based calculations. Efficient
grid-based approaches exist, for example, for the calculation of Bader
charges\cite{Henkelman2006, *Sanville2007, *Tang2009,
  *Yu2011}. Improved partial charge estimation methods such as the
iterative Hirshfeld method\cite{Bultinck2007} have also been recently
made available for condensed matter calculations \cite{Vanpoucke2013,
  *Zicovich-Wilson2016}. However, due to the robustness of the
generalized PM scheme, the use of such accurate partial charges is
unnecessary as equally localized orbitals are obtainable with simpler
charge estimates\cite{Lehtola2014}.

In this article, we describe the theory and implementation of the
generalized PM method for the formation of Wannier functions as a step
to introduce this powerful approach for the chemical interpretation of
electronic structure calculations to the toolbox of the condensed
matter scientist. Our PMWF implementation supports both of the
aforementioned, commonly used representations of the wave functions
for solid state calculations -- PWs and FD grids -- as well as the
linear combination of atomic orbitals (LCAO) approach. In addition,
$\bk$-point sampling is supported in our implementation. We find that
the well-known superiority of the PM scheme to the FB scheme in the
preservation of $\sigma$ and $\pi$ bonds is carried over to the
Wannier functions: PMWFs turn out to be pure $\sigma$ or $\pi$
orbitals, while FBWFs in most cases have mixed $\sigma$ and $\pi$
character. Furthermore, PMWFs are found to be as localized as FBWFs in
all of the systems studied.

The article is organized as follows. The generalized PM scheme of
\citeref{Lehtola2014} is briefly reviewed in the following Method
section. Then, in the Implementation section, the specifics of
extending the generalized PM scheme to the formation of Wannier
functions are presented, and the two weight function schemes used in
the present work (Hirshfeld and Wigner--Seitz partitioning) are
described. In the Computational Details section, we describe how the
calculations on various systems were performed, after which the
calculated results for several systems are presented in the Results
section.  The article concludes with a Summary and Discussion section.


\section{Method}

The generalized PM method is briefly reviewed below. A more detailed
account is given in \citeref{Lehtola2014}. The localization is done
separately for the orbitals in each spin block, and in the following
we will omit the spin indices. Localization is achieved by maximizing
\begin{equation}\label{eq:pm}
  \mathcal{P}(\bW) = \sum_n^{N_\mr{occ}}
  \sum_{\mca=1}^{N_\mca}\left[Q_{nn}^{\mca}(\bW)\right]^p
\end{equation}
where $n$ sums over the $N_\mr{occ}$ occupied states, $\mca$ sums over
all the $N_\mca$ atoms in the system, $p$ is a penalty exponent ($p=2$
in the present work), and $\bW$ is a unitary matrix that connects the
COs $\phi_S$ to the LOs $\psi_n$ as
\begin{align}
  \psi_n(\br) =&\ \sum_R W_{Rn}\phi_R(\br) \label{eq:loco} \\
  \phi_S(\br) =&\ \sum_m W^*_{Sm}\psi_m(\br)
\end{align}
The gradient descent procedure used for the optimization of \eqref{pm}
with respect to $\bW$ has been described elsewhere\cite{Lehtola2013a,
  Lehtola2014}. An analogous problem of optimizing complex-valued
occupied-occupied rotations also arises within self-interaction
corrected density functional theory calculations\cite{Lehtola2014a},
for which another parametrization of the optimization problem,
including optional stability analysis, has been presented in
\citeref{Lehtola2016}. Instead of simultaneous global orbital
rotations, the optimization could also be performed using $2 \times 2$
Jacobi rotations of orbital pairs\cite{Edmiston1963, Pipek1989}. While
this approach generally works well, it may converge onto different
solutions including saddle points depending on the orbital ordering,
thus we prefer the global approach mentioned above. This choice is
furthermore justified by the use of periodic boundary conditions that
necessitate the use of complex orbitals for ${\bf k} \neq {\bf 0}$.
We are not aware of previous discussions of complex Jacobi
rotations. Note that it might also be possible to circumvent the need
for orbital optimization in the localization: for instance, a
non-iterative variant of the real-space PM methods discussed in
\citeref{Lehtola2014} has recently been suggested by
He{\ss}elman\cite{Hesselmann2016}, the extension of which to the case
of periodic boundary conditions could be studied in future work.

The quantity $Q^\mca_{mn}$ that appears in \eqref{pm} is the atomic
partial charge matrix for atom $\mca$ in the LO basis, which has the
properties
\begin{align}
  Q^\mca_{nn}\geq&\ 0, \label{eq:nonneg} \\
  \sum^{N_\mca}_\mca Q^\mca_{nn} =&\ 1, \label{eq:norm}
\end{align}
\eqref{nonneg} stating that $Q^\mca_{mn}$ must represent a
non-negative norm and \eqref{norm} that the LOs are normalized. Then,
the number of electrons localized on atom $\mca$ can be obtained as
\begin{align}
  \sum_n^{N_\mr{occ}}Q^\mca_{nn} =&\ N_\mr{el}^\mca, \label{eq:pca}
\end{align}
by summing over the individual orbital contributions. The (total)
partial charge $q^\mca$ on atom $\mca$ can be calculated as
\begin{equation} \label{eq:atomic-q}
  q^\mca = Z^\mca - N_\mr{el}^\mca,
\end{equation}
where $Z^\mca$ is the atomic number of atom $\mca$; if
pseudopotentials are used, $Z^\mca$ is the effective atomic number
without the frozen core electrons.

While a variety of estimates for the atomic charges can be
used,\cite{Lehtola2014} we focus here exclusively on real-space
methods. The total electron density, $n(\br)$, is partitioned in
real-space into atomic densities through the atomic weight function
$w_\mca(\br)$
\begin{equation} \label{eq:atomic-n}
  n_\mca(\br) = w_\mca(\br)n(\br).
\end{equation}
The number of electrons on atom $\mca$ is then obtained as an integral
\begin{equation} \label{eq:atomic-Nel}
  N^\mca_\text{el} = \int n_\mca(\br){\rm d}^3r.
\end{equation}
The necessary criteria for $w_\mca$ are
\begin{equation} \label{eq:w-crit1}
  0 \leq w_\mca(\br) \leq 1
\end{equation}
and
\begin{equation} \label{eq:w-crit2}
  \sum_\mca^{N_\mca}w_\mca(\br) = 1,
\end{equation}
which allow for a great deal of freedom when choosing the weight
functions. In this formulation, $Q^\mca_{mn}$ is obtained as
\begin{equation}\label{eq:pcm}
  Q^\mca_{mn} = \int \psi_m^*(\br)w_\mca(\br)\psi_n(\br) {\rm d}^3r,
\end{equation}
where $\psi_m$ and $\psi_n$ are LOs, and $w_\mca(\br)$ is a real-space
weight function corresponding to atom $\mca$. Again, since various
weight functions for the generalized PM method have previously been
found to give similar localized orbitals,\cite{Lehtola2014} we choose
here simple and efficient methods for constructing the weight
function, as described below in the Implementation section.

The reason why the generalized PM and PMWF methods do not mix $\sigma$
and $\pi$ bonds in planar systems lies in the orbital rotation
gradient for orbitals $i$ and $j$ which is proportional
to\cite{Lehtola2013a} $Q_{ij}^\mca$. The rotation gradient must vanish
when the optimization objective function has been maximized. However,
$Q_{\sigma \pi}^\mathcal{A}=0$ for a real-space division of the
molecule, assuming the the weight function is even under reflection
over the molecular plane.\cite{Lehtola2014}. Thus, even if the
optimization is started from orbitals that have mixed $\sigma$-$\pi$
character, the orbital rotation gradient will end up separating these
components into $\sigma$- and $\pi$-type orbitals. A similar
discussion including stability analysis has been given in
\citeref{Pipek1989} for the original Pipek--Mezey formulation using
Mulliken charges.

The corresponding matrix elements in the FB optimization, however, do
not possess similar symmetry properties and as a result FB (and FBWF)
end up mixing $\sigma$ and $\pi$ orbitals. FBWF can be coaxed to
reproduce $\sigma$-$\pi$ separation if one allows mixing of occupied
and unoccupied orbitals\cite{Thygesen2005, Thygesen2005a}. However,
the orbitals then become partially occupied instead of being fully
occupied, or not occupied at all. Therefore, the interpretation of the
results becomes more challenging, as a single electron may be
simultaneously represented by multiple orbitals. As an extreme, if all
valence unoccupied orbitals are included in such a procedure, the
localization will just reproduce basis functions localized on the
atoms and the resulting orbitals end up being devoid of chemical
information about the system being studied. In contrast, the PM and
PMWF approaches yield proper $\sigma$-$\pi$ separation without needing
to resort to partial occupation numbers, which makes e.g. the chemical
interpretation and use in post-Hartree--Fock methods simpler.

Alternatively, when the global symmetry of the system allows for the
\emph{a priori} identification of $\sigma$ and $\pi$ states,
localization procedures can be restricted to operate within the set of
orbitals of a given symmetry. While this procedure enables any
localization method to yield pure $\sigma$ and $\pi$ orbitals, the
constraint of no mixing of $\sigma$ and $\pi$ states in the
optimization also means that the resulting LOs may be less localized
than what they would be without such constraints. An application of
symmetry restrictions to FBWF has been presented
recently\cite{Sakuma2013}. However, point group symmetries that allow
for the \emph{a priori} grouping of orbitals are only present in
limited special cases.  The PMWF approach presented here is a general
purpose tool where special restrictions do not need to be introduced
to obtain proper separation of $\sigma$ and $\pi$ orbitals.


\section{Implementation}

Because the maximization of \eqref{pm} needs many evaluations of
$Q^\mca_{mn}$, a $N_\mr{occ}\times N_\mr{occ}$ matrix for
every atom $\mca$, it makes technically more sense to evaluate
\begin{equation}\label{eq:copcm}
  \tilde{Q}^\mca_{RS} = \int \phi_R^*(\br)w_\mca(\br)\phi_S(\br) {\rm d}^3r,
\end{equation}
where $\tilde{Q}^\mca$ is the atomic partial charge matrix in the CO
basis and $\phi_R$ and $\phi_S$ are occupied COs, to construct
$Q^\mca_{mn}$ as
\begin{equation}\label{eq:rot}
  Q^\mca_{mn} = \sum_{RS}W^*_{Rm}\tilde{Q}^\mca_{RS}W_{Sn}
\end{equation}
where $R$ and $S$ index the occupied COs. This way, the expensive
numerical integration in \eqref{copcm} needs to be done only once at
the beginning of the calculation, and the matrices may be stored on
disk so that only the matrix corresponding to a given atom $\mca$
needs to be kept in memory at a time.

Moving on to a periodic system, the maximization of the objective
function in \eqref{pm} now includes periodic images through the
overlap matrices which are defined in terms of\cite{Resta1999}
primitive lattice vectors $\{G_\alpha\}$ (three for orthorhombic cells
as implemented here) with corresponding weights $\{g_\alpha\}$. The
definitions of the vectors and weights can be found in
\citerefs{Silvestrelli1999, Berghold2000}, which present a
generalization of the overlap matrices for cubic periodic
systems\cite{Silvestrelli1998} to any cell symmetry but restricted to
the $\Gamma$-point.  However, by defining a $\bk$-point
mesh\cite{Thygesen2005a} in terms of an artificially expanded
unit-cell, the method is applicable to periodic systems with
$\bk$-point sampling. A brief overview of the lattice vector
definitions is presented in the appendix.

For periodic systems the COs are represented in terms of Bloch
functions
\begin{equation}
  \phi_{S\bk}(\br) = e^{i\bk\cdot\br}u_{S\bk}(\br)
\end{equation}
where $u_{S\bk}$ has the periodicity of the lattice. In analogy to
\eqref{loco}, the $n$:th LO is given by
\begin{equation}\label{eq:pbcloco}
  \psi_{n,c}(\br) = \frac{1}{\sqrt{N_\bk}}\sum_\bk\sum_SW_{Sn}^\bk
  e^{-i\bk\cdot {\bf R}_c}\phi_{S\bk}(\br)
\end{equation}
relative to unit cell $c$, where $N_\bk$ is the number of $\bk$-points
and ${\bf R}_c$ is any Bravais lattice vector. The objective function
of \eqref{pm} becomes
\begin{equation}\label{eq:pbcpm}
  \mathcal{P}(\bW)=\sum_n^{N_\mr{occ}}\sum_\mca^{N_\mca}\sum_\alpha^{N_\alpha}
  g_\alpha\!\mid\!Q^\mca_{\alpha,nn}(\bW)\!\mid^p,
\end{equation}
where $\mathbf{\tilde{Q}}^\mca$ and $\mathbf{W}$ now take on a
\bk-dependent form
\begin{equation}\label{eq:pbcrot}
  Q^\mca_{\alpha,mn} = \sum_{\bk\bk'}\sum_{RS}
  \left(W^{\bk}_{Rm}\right)^* \tilde{Q}^{\mca,\bk\bk'}_{\alpha,RS}W^{\bk'}_{Sn}.
\end{equation}
  \Eqref{copcm} can be written as\cite{Resta1999}
\begin{equation}\label{eq:pbccopcm}
  \tilde{Q}^{\mca,\bk\bk'}_{\alpha,RS} =
  \int u^*_{R\bk}(\br)w^\mca(\br)u_{S\bk'}(\br)e^{i(\bk'-\bk-\mathbf{G}_\alpha)\cdot\br}{\rm d}^3r
\end{equation}
which is nonzero only when $\bk'=\bk+\bG_\alpha$. This reduces the
double sum over $\bk$ and $\bk'$ in \eqref{pbcrot} to a single sum
over $\bk$.  However, the optimization of the unitary rotation
matrices at the \bk-points, $\{{\bf W^k}\}$, depends on neighbouring
\bk-points as well ($\bk'=\bk+\bG_\alpha$).

As in typical implementations\cite{Boys1966} of its parent
method
-- the FB approach\cite{Foster1960} -- the
objective function optimized in practical implementations of the FBWF
method\cite{Marzari1997, Souza2001, Thygesen2005, Thygesen2005a,
  Marzari2012} is not based on the minimization of the orbital spread,
but on the equivalent task of maximizing the sum of squares of
distances of orbital centroids from the origin of the coordinate
system\cite{Silvestrelli1999, Resta1999}
\begin{equation}\label{eq:FBWF-crit}
  \mathcal{L} (\bW) = \sum_n^{N_\mr{occ}}\sum_\alpha^{N_\alpha}g_\alpha\!\mid\!Z^\alpha_{nn}\!\mid^2
\end{equation}
where
\begin{equation}\label{eq:FBWF-Znn}
  Z^\alpha_{nn} = \sum_{RS}W^*_{Rn}Z^\alpha_{SR}W_{Sn}
\end{equation}
and
\begin{equation}\label{eq:FBWF-ZSR}
  Z^\alpha_{SR} = \langle\phi^*_R\!\mid\!e^{-i\bG_\alpha\cdot\br}\!\mid\!\phi_S\rangle,
\end{equation}
and a similar unitary optimization problem is solved as in the
PMWF method.

The task of maximizing the objective function of \eqref{pbcpm}, when restricted to $p=2$, is similar to the task of maximizing \eqref{FBWF-crit}, so
methods used to maximize
\eqref{FBWF-crit} -- employing, for example, steepest descent
algorithms\cite{Thygesen2005a, Silvestrelli1998} -- are applicable
to \eqref{pbcpm}.
This means that existing FBWF codes based on maximizing the objective
function of the \eqref{FBWF-crit} form could easily be modified to generate
PMWFs instead.


\subsection{Weight functions}

Two different forms of the weight function were chosen for the present
work: a Gaussian weight function, which results in a fuzzy real-space
partitioning of the system into atomic regions, and Wigner--Seitz
partitioning, which divides the system into non-overlapping atomic
regions. In agreement with the results of \citeref{Lehtola2014}, these
two qualitatively different weight functions are found to produce
similar LOs despite predicting strikingly different atomic partial
charges, as will be seen in the Results section.

\subsection{Gaussian weight}

A simple choice of the weight function (\eqref{w-crit1,w-crit2}) is
obtained using a Hirshfeld-type\cite{Hirshfeld1977} partitioning
\begin{equation}\label{eq:hirsh}
  w_\mca(\br) = \frac{\overline{n}_\mca(\br)}
  {\sum_{\mca'=1}^{N_\mca}\overline{n}_{\mca'}(\br)}
\end{equation}
where $\overline{n}_\mca$ are spherically symmetric functions that
could be gas-phase atom densities (as in the original Hirshfeld
scheme), or iteratively defined atom densities such as in the
iterative Hirshfeld\cite{Bultinck2007, Bultinck2007a, Verstraelen2013}
or the iterative Stockholder schemes\cite{Lillestolen2008,
  Lillestolen2009}. Based on the experience from \citeref{Lehtola2014}
that the generalized PM scheme is insensitive to the partial charge
estimate, we choose simple Gaussian model densities
\begin{equation}\label{eq:gauss}
  \overline{n}_\mca(\br) = \frac{N_{\mr{el},\mca}}{\gamma_\mca\sqrt{2\pi}}
  \exp\left\{-\frac{(\br - {\bf R}^\mca)^2}
                   {2\gamma_\mca^2}\right\}
\end{equation}
where $N_{\mr{el},\mca}$ is the (effective) number of electrons on
atom $\mca$ and ${\bf R}^\mca$ is its position. This model was used by
Oberhofer and Blumberger\cite{Oberhofer2009} in the context of
constrained density functional theory. They found that qualitatively
similar results were obtained when using spherical neutral
free atom densities and the Gaussian model densities. Furthermore, the
charge constrained energy did not depend strongly on the choice of
decay parameter $\gamma_\mca$ in the range $\gamma_\mca\in [0.5
 \text{ \AA}, 1.0 \text{ \AA}]$.

Results presented here using the Gaussian model densities
were obtained with a decay parameter value of $\gamma_\mca = 0.5$ \AA\ for
all types of atoms, unless stated otherwise. A different choice for
the decay parameter results in practically the same localized
orbitals, but with different total partial atomic charges
(\eqref{pca}).  The form of \eqref{gauss} is
convenient as the fast decay of the model density allows for
efficient spatial screening of contributions.

\subsection{Wigner--Seitz partitioning}

Alternatively, the weight function
can be based on non-overlapping atomic regions defined using, for example the
Bader\cite{Bader1990} or the Wigner--Seitz schemes, for which the
weight function is a step function
\begin{equation}\label{eq:wsfunc}
  w_\mca(\br) = \begin{cases}
    1 & \quad \text{if $\br \in \mca$} \\
    0 & \quad \text{otherwise}
  \end{cases}
\end{equation}
that clearly satisfies the criteria of \eqref{w-crit1, w-crit2}. The
Wigner--Seitz scheme corresponds to the Voronoi scheme discussed in
\citeref{Lehtola2014} since Wigner--Seitz cells are Voronoi cells.
The Wigner--Seitz scheme is parameter-free, easy to construct on a
grid, and forms the second class of weight functions chosen for the
present work. Furthermore, this scheme is even more convenient than
the Gaussian weights for use with periodic systems, since it has a
clear cut-off in any dimension which is particularly convenient when
periodicity is present.


\section{Computational Details}

The electronic structure calculations in this work are performed with
the GPAW program\cite{Mortensen2005a, *Enkovaara2010}. There, the wave
functions can be represented using any of the three aforementioned
approaches: with real-space grids, atomic orbitals\cite{Larsen2009},
or plane waves, and any of these representations can be used in the
present PMWF implementation.  The present results have been obtained
using the FD approach, unless otherwise stated. We also present a few
cases where the PW and LCAO approaches have been used.

All the electronic structure calculations in this work made use of the
PBE\cite{Perdew1996, *Perdew1997} exchange-correlation functional,
using a 360 eV kinetic energy cutoff for PWs, a 0.18 \AA\ spacing for
FD, and the "dzp"-basis for the LCAO. A convergence criterion of 0.05
eV/\AA\ was used for the forces to relax the nuclear degrees of
freedom. Core electrons were described with the projector augmented
wave (PAW) method\cite{Blochl1994}. Periodic boundary conditions
employed a Monkhorst--Pack grid\cite{Monkhorst1976} to sample the
Brillouin zone with dimensions of (3,1,1), (3,3,1) and (3,3,3) for the
systems with periodicity in one, two, and three dimensions,
respectively. The unit cell dimension was relaxed along the directions
on which periodic boundary conditions were applied, whereas a 7
\AA\ vacuum region was included on both sides of the system along
non-periodic directions.

The PMWF method has been implemented as a standalone object-oriented
package in the Atomic Simulation Environment\cite{Bahn2002} (ASE)
library. The implementation supports the use of PAW for the core
electrons; the PAW specific details of the implementation are given in
the Appendix. While our results were obtained with GPAW, the
implementation in ASE can also be used with other software
packages for which ASE has an interface, such as
ABINIT\cite{Gonze2002, *Gonze2009}, NWCHEM\cite{Valiev2010}, and
VASP\cite{Kresse1996a, *Kresse1999}.

All surfaces are drawn with the open-source software package
Jmol\cite{Jmol}, at an isosurface value corresponding to a 75\%
density cut-off as described in the appendix of
\citeref{Lehtola2014}. In contrast to the common approach of using a
fixed isosurface value for all orbitals, the density isosurfaces
constitute an unambiguous visualization method, as the value of the
isosurface will reflect how localized an orbital is. This way all
orbitals are treated on equal footing, regardless of their character.


\subsection{Analysis of localized states}


To analyze the $\sigma$- or $\pi$-bond mixing of the PMWFs and FBWFs,
the COs are first classified into $\sigma$ and $\pi$ states. This
operation is trivial for linear and planar systems, as the mirror
symmetry operator ($\mathcal{M}$) through the molecular or periodic
plane leaves $\sigma$ states unchanged whereas $\pi$ states undergo a
phase change; that is
\begin{align}
  \mathcal{M}\sigma(x,y,z) \equiv \sigma(x,y,-z) = \sigma(x,y,z) \label{eq:sym-sigma} \\
  \mathcal{M}\pi(x,y,z) \equiv \pi(x,y,-z) = -\pi(x,y,z) \label{eq:sym-pi}
\end{align}
The $\sigma$- and $\pi$-bond mixing of the PMWFs and FBWFs can then
easily be calculated, since the expansion coefficients of the $n$:th
PMWF or FBWF in terms of the COs are the $n$:th rows of the optimized
unitary matrix (\eqref{pbcloco}). The fractions of $\sigma$ and $\pi$
type character in the $n$:th LO, $f_n^\sigma$ and $f_n^\pi$,
respectively, can be obtained as
\begin{align}
  f_n^\sigma =& \sum_{\sigma\text{-type } S}W_{Sn}^* W_{Sn} \label{eq:f-sigma} \\
  f_n^\pi    =& \sum_{\pi\text{-type } S}W_{Sn}^* W_{Sn} \label{eq:f-pi}
\end{align}
Because there is no unique way of defining orbital locality, the
values of the PMWF and FBWF objective functions, \eqref{pbcpm,
  FBWF-crit}, are used as measures of localization for both PMWF and
FBWF orbitals. The FBWF measures are computed with the pre-existing
implementation in ASE (see \citeref{Thygesen2005a}), and the PMWF
measures are obtained with the method described here.


\section{Results}

A variety of systems were calculated with KS-DFT as described above in
the Computational Details section, and the resulting COs localized
using both the FBWF and PMWF approaches. COs and LOs for various
systems are shown in \figrangeref{boron_nitride}{benzene_crystal}.  In
some cases the PMWF and FBWF methods give qualitatively similar
results and the visual comparison is, hence, omitted, but the results
are noted in the main text. The systems studied range from isolated
molecules (benzene, coronene, supercoronene), to systems with one-
(polyethylene, polyacetylene, carbyne, armchair nanoribbons), two-
(graphene, boron nitride) or three-dimensional (benzene crystal)
periodicity.


\subsection{Comparison of basis and partitioning functions}

Unlike the mathematically ill-founded Mulliken charges that were used
in the original Pipek--Mezey scheme\cite{Pipek1989}, the generalized
Pipek--Mezey schemes\cite{Lehtola2014} of \citerefs{Cioslowski1991,
  Alcoba2006, Lehtola2014, Knizia2013} rely on mathematically
well-defined partial charge estimates that provide smooth convergence
to the basis set limit. \Tabref{basiscomparison} presents the
generalized Pipek--Mezey objective function values and atomic charge
estimates for cis-polyacetylene, where the COs are described with the
FD, PW, or LCAO bases in GPAW. A convergence test in the FD and PW
bases reveals that the value of the objective function (\eqref{pbcpm})
and the partial atomic charges (\eqref{atomic-q}) reach constant
values at a grid-spacing of 0.3 \AA{} or a kinetic energy cut-off of
300 eV, for the two bases, respectively.  These are well within the
typical range of values used for FD and PW calculations, and should
not pose problems for typical applications.

\begin{table}
  \centering
  \caption{Value of the objective function, \eqref{pbcpm}, and atomic
    charge estimates, \eqref{atomic-q}, for cis-polyacetylene (see main text).}
      \begin{tabular*}{\columnwidth}{@{\extracolsep{\fill}} l | c | c }
    \hline
    basis & $\mathcal{P}(\mathbf{W})$ & $Q_\mr{C}$ / $-Q_\mr{H}$ \\
    \hline
    LCAO & 8.74 & 0.058 \\
    FD   & 8.71 & 0.062 \\
    PW   & 8.70 & 0.065 \\
    \hline
  \end{tabular*}\label{tab:basiscomparison}
\end{table}

In the PMWF method the atomic partial charge matrix, \eqref{copcm},
depends on the choice of the atomic weight function. This dependence
suggests that the orbitals that maximize the PMWF objective function
could also depend on the weight function used. However, in
\citeref{Lehtola2014} the orbitals were found by visual comparison to
be remarkably insensitive to the choice of the partial charge estimate,
even while the resulting atomic (total) partial charge assignment
varied greatly. We confirm this result for the partitioning functions
used in the present work -- the Hirshfeld-type weight function using
Gaussian model densities (\eqref{hirsh,gauss}, respectively) with
various choices for the atomic decay parameters, as well as the
Wigner--Seitz weight function (\eqref{wsfunc}).

The number of electrons localized on the atoms (\eqref{atomic-Nel}) in
a periodic boron nitride sheet was estimated with the two choices of
weight functions: The Wigner--Seitz (WS) function (\eqref{wsfunc}),
and the Hirshfeld-type (H) function (\eqref{hirsh}). Different choices
for the decay parameter in the Gaussian model density (\eqref{gauss})
for boron $\gamma_\mr{B}$ were considered, while the decay parameter
for nitrogen is kept fixed at $\gamma_\mr{N} = 0.5$
\AA{}. \Tabref{partialcharges} presents the resulting atomic charge
estimates (\eqref{pca}), where the partial charge on the nitrogen
atoms is the negative of that on the boron atoms.  The Wigner--Seitz
and Hirshfeld function with symmetric decay factors describe boron as
a donor, while increasing the value of the decay parameter of boron
(thus ascribing more space -- and electrons -- to it) makes it an
acceptor in the partial charge analysis.

\begin{figure}
  \centering
  \includegraphics[width=.49\textwidth]{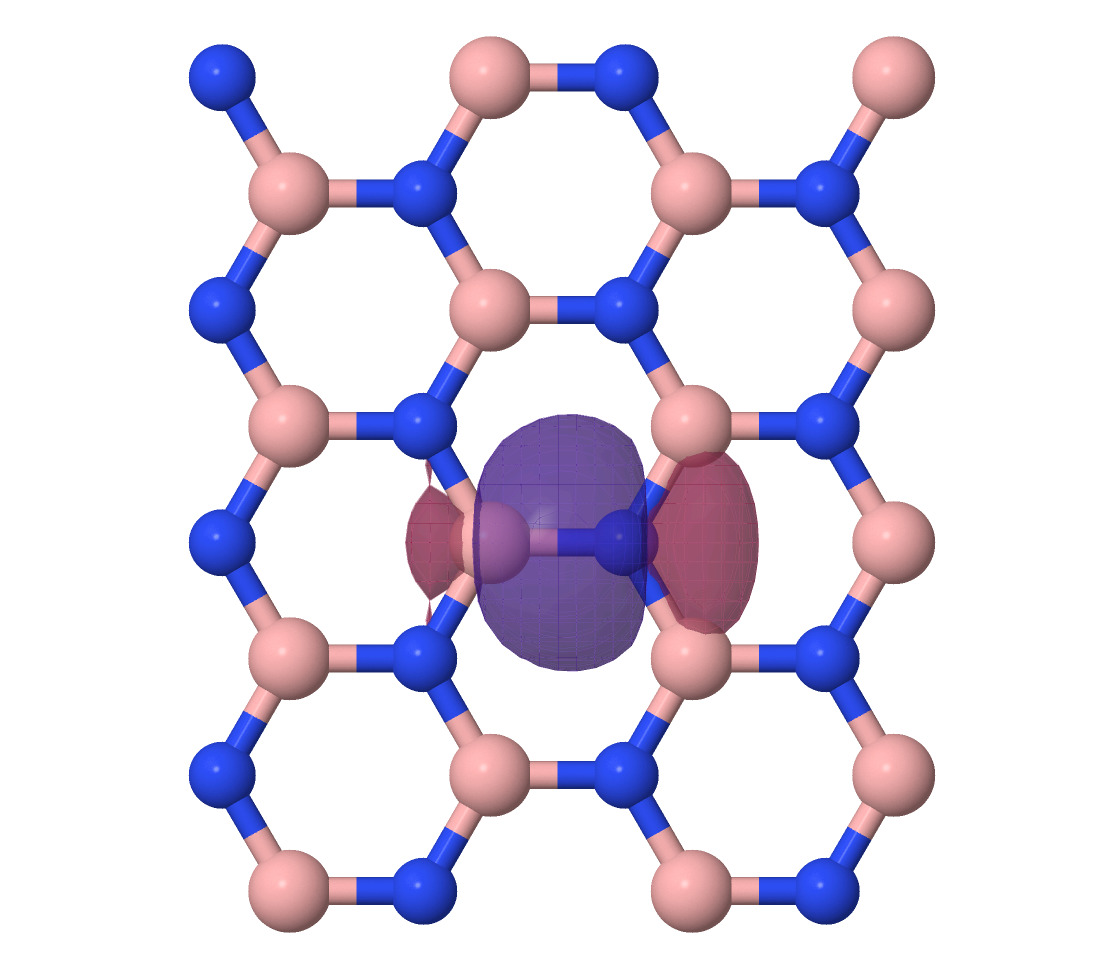}
  \hfill
  \includegraphics[width=.49\textwidth]{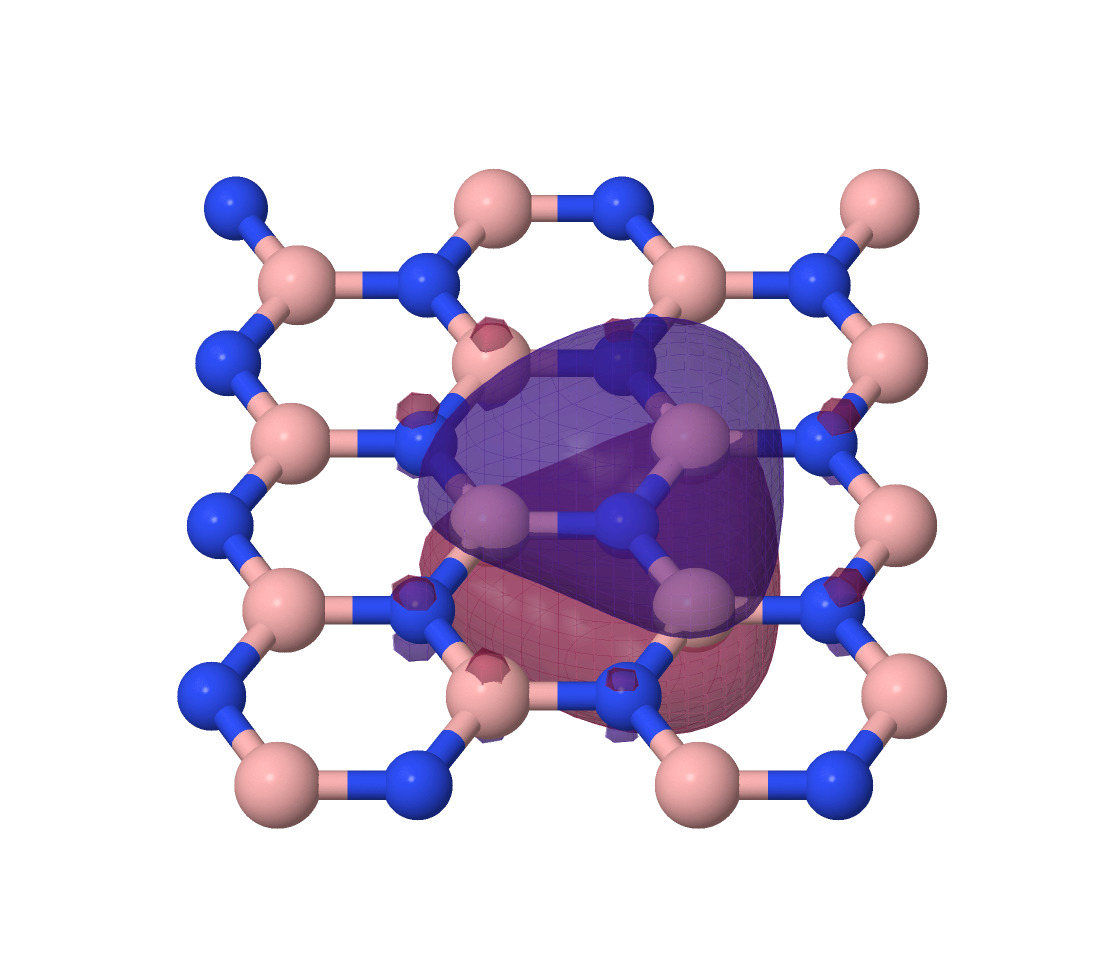}
  \caption{Pipek--Mezey orbitals in boron nitride, represented by
    B$_{16}$N$_{16}$ sheet in a two-dimensional periodic lattice,
    obtained with the Wigner--Seitz weight function (see main text).}
  \label{fig:boron_nitride}
\end{figure}

While the partial charges on the atoms estimated by the various models
are clearly different -- showing a variation larger than one
elementary charge unit -- they all result in nearly identical LOs, and
no discernible difference can be seen for the 75\%
density\cite{Lehtola2014} isosurfaces. For this reason, only the PMWFs
for the Wigner--Seitz weight scheme are shown in \figref{boron_nitride}.
This is in agreement with the result found for molecules in
\citeref{Lehtola2014}, highlighting the versatility of the PM method
in achieving orbital localization even with partial charge estimates
that disagree with chemical intuition -- as long as the charge
estimates are mathematically well-defined\cite{Lehtola2014}.

\begin{table}
  \centering
  \caption{Atomic charge estimates, \eqref{atomic-q}, in a boron
    nitride sheet using Wigner--Seitz (WS) and Hirshfeld (H) function
    definition of partial charges with three different choices for the
    decay parameter of the boron atoms, $\gamma_\mr{B}$ (see main text). }
  \begin{tabular*}{\columnwidth}{@{\extracolsep{\fill}} l | c }
    \hline
    $w_\mca(\br)$ & $Q_\mr{B}$ / $-Q_\mr{N}$ \\
    \hline
    WS & 0.369 \\
    H, $\gamma_\mr{B}=0.50$ \AA & 0.490 \\
    H, $\gamma_\mr{B}=0.75$ \AA & -0.224 \\
    H, $\gamma_\mr{B}=1.00$ \AA & -0.651 \\
    \hline
  \end{tabular*}\label{tab:partialcharges}
\end{table}

To quantify the similarity of the PMWF orbitals obtained with the
different choices of the atomic charge definitions, their overlap was
evaluated.  For two sets of PMWFs, $\{\psi_n^A\}$ and $\{\psi_n^B\}$,
obtained from the use of two different weight functions $w^A$ and
$w^B$ in the localization procedure, the residual overlap matrix is
evaluated
\begin{equation} \label{eq:diff-overlap}
  R^\text{AB}_{mn} = \left| \braket{\psi_m^A | \psi_n^B} \right|^2 - \delta_{mn},
\end{equation}
in analogy to what was recently used to analyze the orbital
convergence of self-interaction corrected density functional theory
calculations\cite{Lehtola2016}. The ordering and complex
phases of the orbitals for two identical calculations started from a
different random starting point may differ\cite{Lehtola2016}
(e.g. $\ket{\psi_n}$ spans the same orbital density as
$\ket{\psi_n^*}$ while their overlaps differ), but we circumvent both
of these problems by carrying out one optimization for the PMWFs with one
choice for the weight function -- the Hirshfeld function with symmetric
decay factors -- and then by using the orbitals from
this calculation as starting guesses for the other choices of the weight function.

The diagonal of ${\bf R}^\text{AB}$ measures the similarity of the
localized orbitals produced using weight schemes A and B. Thus,
the similarity is quantified by
\begin{eqnarray}
  R^\text{AB}_\text{max} & = \max_{n} R^\text{AB}_{nn} \label{eq:R-max}, \\
  R^\text{AB}_\text{rms} & = \sqrt{ \sum_{n=1}^N (R^\text{AB}_{nn})^2 / N }, \label{eq:R-rms}
\end{eqnarray}
that is, the maximum and root-mean-square (rms) deviations,
respectively, from perfect insensitivity to the used weight
metric. The results are shown in \tabref{orbitaloverlap}, from which
the degree of similarity is apparent. The overlaps are close to
unity, once again confirming the robustness of the generalized
Pipek--Mezey scheme. In the rest of the calculations presented here, the
Hirshfeld-type weighting factor was used with equal-valued decay factors.

\begin{table*}
  \centering
  \caption{Analysis of the similarity of PMWFs obtained using
    various atomic weighing schemes. The first row and first column
    indicate $A$ and $B$ in \eqref{diff-overlap}. The lower triangle
    gives $\text{lg}(R_\text{rms})$ (\eqref{R-rms}) and the upper
    triangle gives $\text{lg}(R_\text{max})$ (\eqref{R-max}).}
  \begin{tabular*}{\textwidth}{@{\extracolsep{\fill}} l | c  c  c  c}
    \hline
    $w_\mca(\br)$ & WS & H, $\gamma_\mr{B}=0.50$ \AA & H,
    $\gamma_\mr{B}=0.75$ \AA & H, $\gamma_\mr{B}=1.00$ \AA \\
    \hline
    WS                          & & -3.1 & -3.1 & -3.0 \\
    H, $\gamma_\mr{B}=0.50$ \AA & -3.3 & & -3.7 & -3.2 \\
    H, $\gamma_\mr{B}=0.75$ \AA & -3.3 & -3.8 & & -3.7 \\
    H, $\gamma_\mr{B}=1.00$ \AA & -3.2 & -3.4 & -4.2 & \\
    \hline
  \end{tabular*}\label{tab:orbitaloverlap}
\end{table*}


\subsection{Localized states and $\sigma$- and $\pi$-bond mixing}

The mixing of $\sigma$- and $\pi$-bond orbitals is analyzed by
calculating the $\sigma$ and $\pi$ projections of the PMWFs and FBWFs
using \eqref{f-sigma, f-pi}, respectively. This analysis is performed
for all planar systems, carbyne, and cis-polyacetylene, and the
results are presented in \tabref{mixing}. Localized states in the set
of PMWFs and FBWFs which are not mixed and clearly represent $\sigma$-
and $\pi$-bonds are denoted $\sigma_{\mca\mca'}$ and
$\pi_{\mca\mca'}$, respectively. Mixed $\sigma$ and $\pi$ states are
denoted $\tau$, following \citeref{Pipek1989}.  It is clear from
\tabref{mixing} that the mixing of $\sigma$- and $\pi$-bond orbitals
in the FBWF method occurs
systematically for the aromatic hydrocarbons. In aromatic systems a
$\pi$-bond orbital exists for every two carbon atoms. The consistent
50/50 mixing of $\sigma$ and $\pi$ states doubles the number of
$\pi$-type orbitals in the FBWF set, and as a result reduces the
number of $\sigma_\mr{CC}$-bond orbitals.  No pure $\pi$-bond orbitals
are found, and a $\tau$ mixed state (``banana'' shaped orbitals, as
coined by Pipek and Mezey\cite{Pipek1989}) exists for every carbon
atom, resulting in a distorted chemical picture for this type of
system.  The set of PMWFs contain a localized $\sigma_\mr{CC}$ for
each possible carbon-carbon bond, and a $\pi_\mr{CC}$ for every two
carbon atoms -- there is no mixing found between the $\sigma$ and
$\pi$ states -- representing the conventional chemical picture
of the aromatic hydrocarbons. Both sets represent all possible
$\sigma_\mr{CH}$-bonds with similar degree of localization.

\begin{table*}
  \centering
  \caption{Number of $\sigma$, $\pi$ and $\tau$ bonding states in the
    set of PMWFs and FBWFs, as well the expansion of the $\tau$ states
    in $\sigma$ and $\pi$ type COs (see main text).} \small
  \begin{tabular*}{1.\textwidth}{@{\extracolsep{\fill}} l | c | c | c}
    \hline
    System          & PMWFs & FBWFs & $\tau$ composition \\
    \hline
    benzene         & 6$\sigma_\mr{CC}$, 6$\sigma_\mr{CH}$, 3$\pi_\mr{CC}$
    & 3$\sigma_\mr{CC}$, 6$\sigma_\mr{CH}$, 6$\tau $
    & $50\%\sigma + 50\%\pi$  \\
    coronene        & 30$\sigma_\mr{CC}$, 12$\sigma_\mr{CH}$, 12$\pi_\mr{CC}$
    & 18$\sigma_\mr{CC}$, 12$\sigma_\mr{CH}$, 24$\tau$
    & $50\%\sigma + 50\%\pi$\\
    supercoronene  & 72$\sigma_\mr{CC}$, 18$\sigma_\mr{CH}$, 27$\pi_\mr{CC}$
    & 45$\sigma_\mr{CC}$, 18$\sigma_\mr{CH}$, 54$\tau$
    & $50\%\sigma + 50\%\pi$ \\
    AC(2,4)         & 40$\sigma_\mr{CC}$, 16$\sigma_\mr{CH}$, 16$\pi_\mr{CC}$
    & 24$\sigma_\mr{CC}$, 16$\sigma_\mr{CH}$, 32$\tau$
    & $50\%\sigma + 50\%\pi$ \\
    AC(3,3)         & 48$\sigma_\mr{CC}$, 12$\sigma_\mr{CH}$, 18$\pi_\mr{CC}$
    & 30$\sigma_\mr{CC}$, 12$\sigma_\mr{CH}$, 36$\tau$
    & $50\%\sigma + 50\%\pi$ \\
    AC(4,3)         & 66$\sigma_\mr{CC}$, 12$\sigma_\mr{CH}$, 24$\pi_\mr{CC}$
    & 42$\sigma_\mr{CC}$, 12$\sigma_\mr{CH}$, 48$\tau$
    & $50\%\sigma + 50\%\pi$ \\
    benzene crystal & 24$\sigma_\mr{CC}$, 24$\sigma_\mr{CH}$, 12$\pi_\mr{CC}$
    & 12$\sigma_\mr{CC}$, 24$\sigma_\mr{CH}$, 34$\tau$
    & $50\%\sigma + 50\%\pi$ \\
    cis-polyacetylene & 8$\sigma_\mr{CC}$, 8$\sigma_\mr{CH}$, 4$\pi_\mr{CC}$
    & 4$\sigma_\mr{CC}$, 8$\sigma_\mr{CH}$, 8$\tau$
    & $50\%\sigma + 50\%\pi$ \\
    carbyne         & 8$\sigma_\mr{CC}$, 8$\pi_\mr{CC}$
    & 4$\sigma_\mr{CC}$, 12$\tau$
    & $33\%\sigma + 66\%\pi$ \\
    graphene        & 48$\sigma_\mr{CC}$, 16$\pi_\mr{CC}$
    & 48$\sigma_\mr{CC}$, 16$\pi_\mr{CC}$
    &  \\
    boron nitride    & 48$\sigma_\mr{BN}$, 16$\pi_\mr{BN}$
    & 48$\tau^A$, 16$\tau^B$
    & $\tau^A: 97\%\sigma + 3\%\pi$ \\
    & & & $\tau^B: 9\%\sigma + 91\%\pi$ \\

    \hline
  \end{tabular*}
  \label{tab:mixing}
\end{table*}

\begin{sidewaysfigure*}
\subfloat[Kohn--Sham\label{fig:nr-KS}]{\includegraphics[width=0.24\textheight]{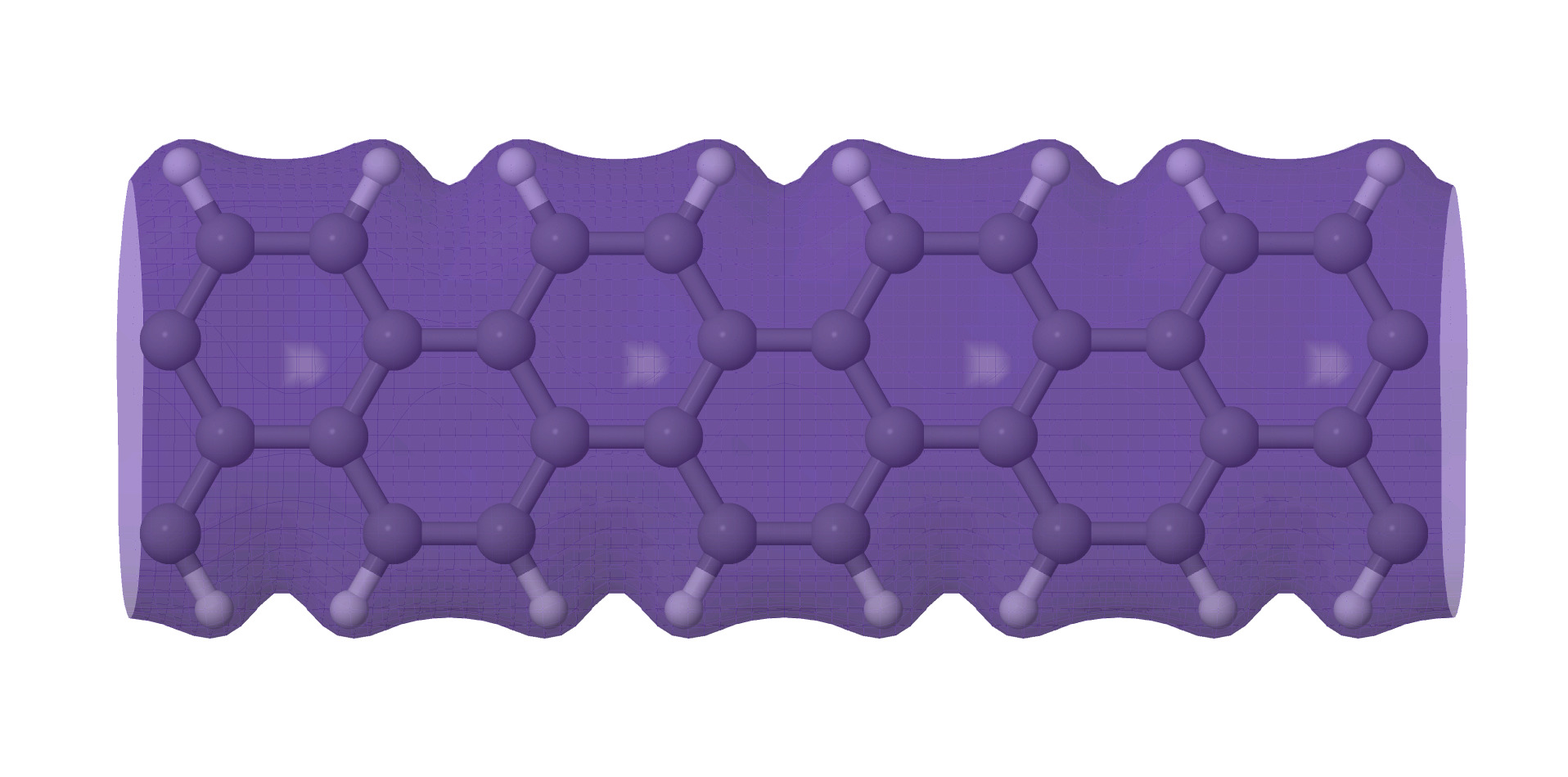}\includegraphics[width=0.24\textheight]{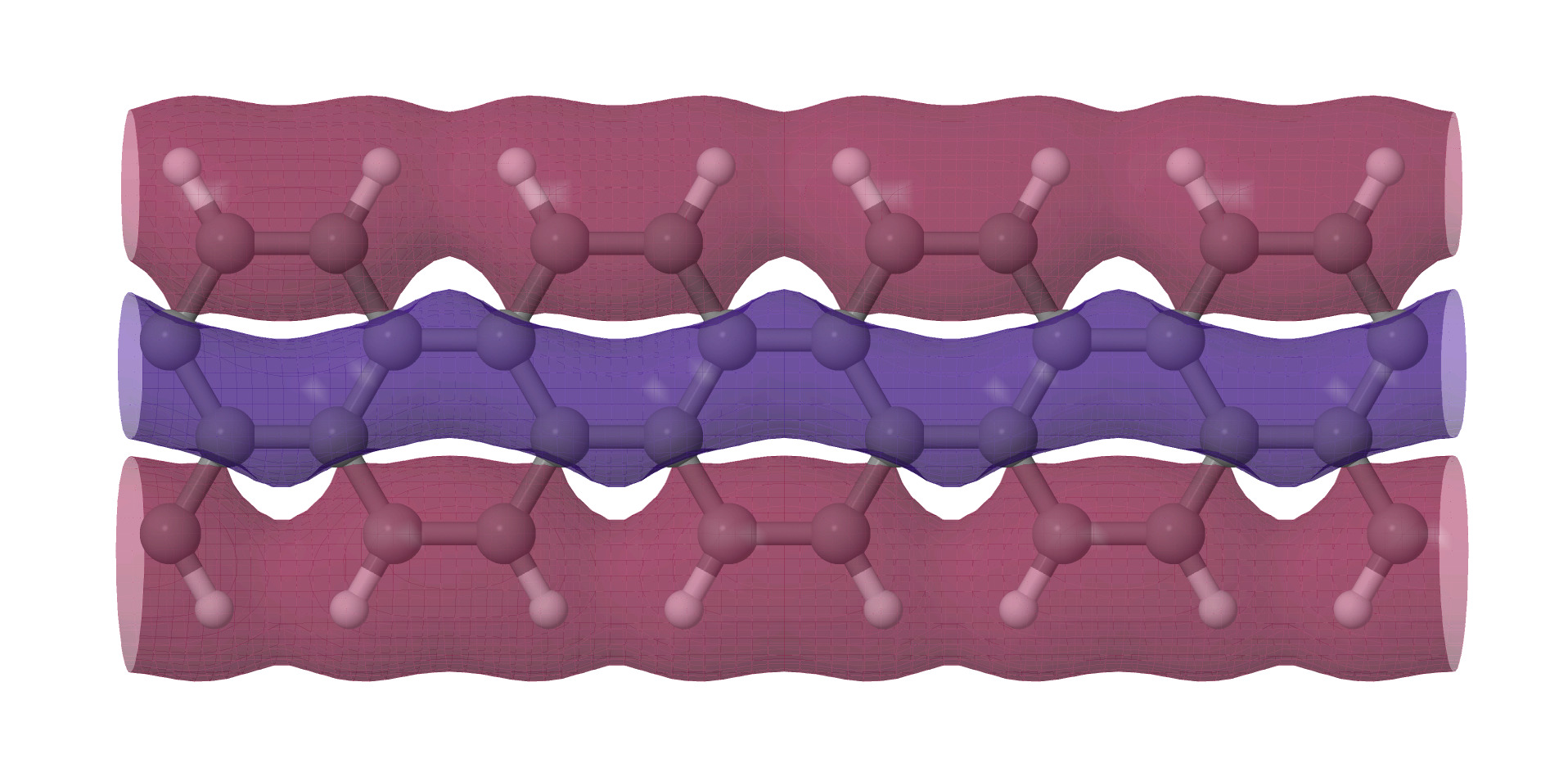}\includegraphics[width=0.24\textheight]{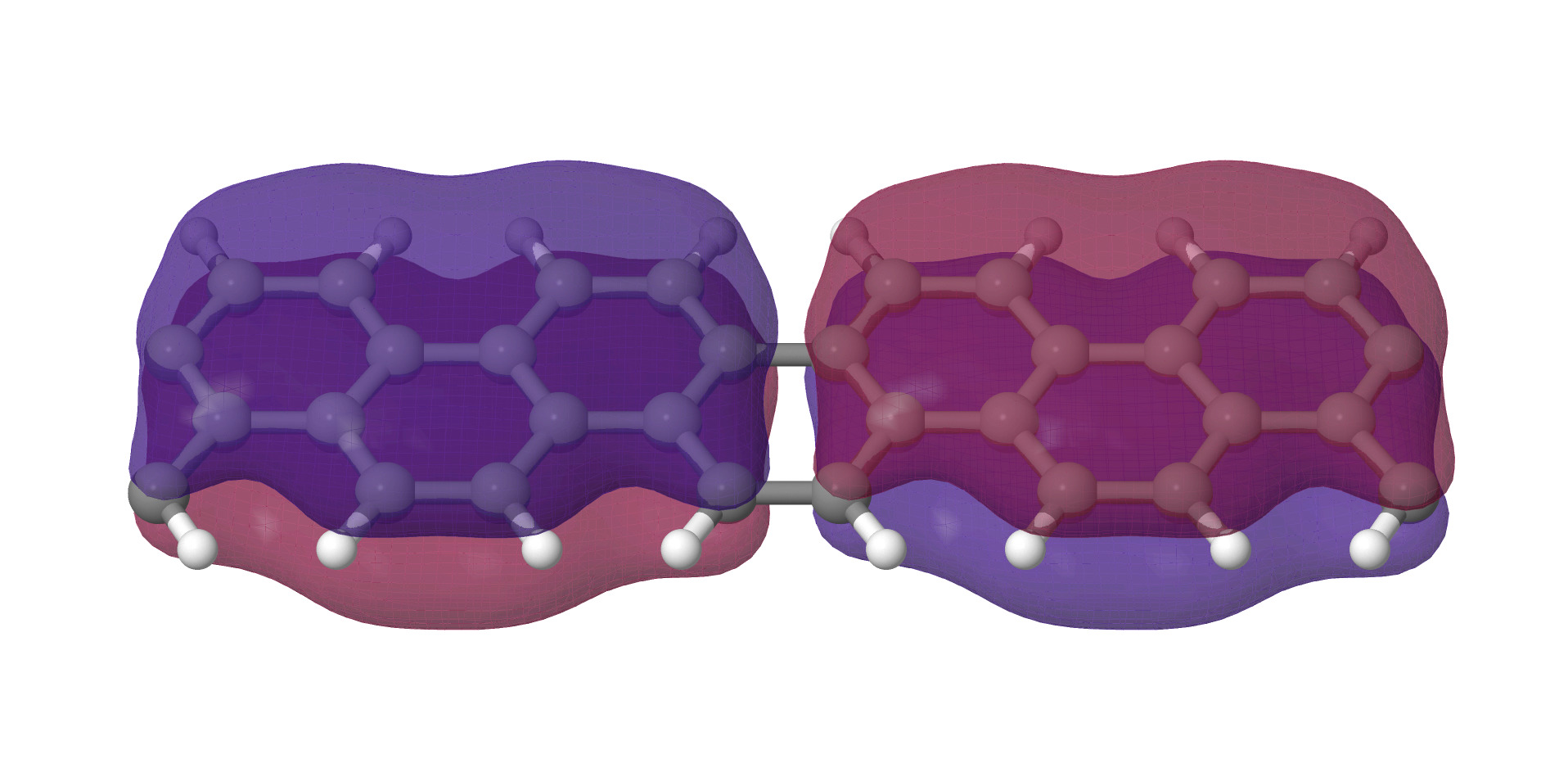}\includegraphics[width=0.24\textheight]{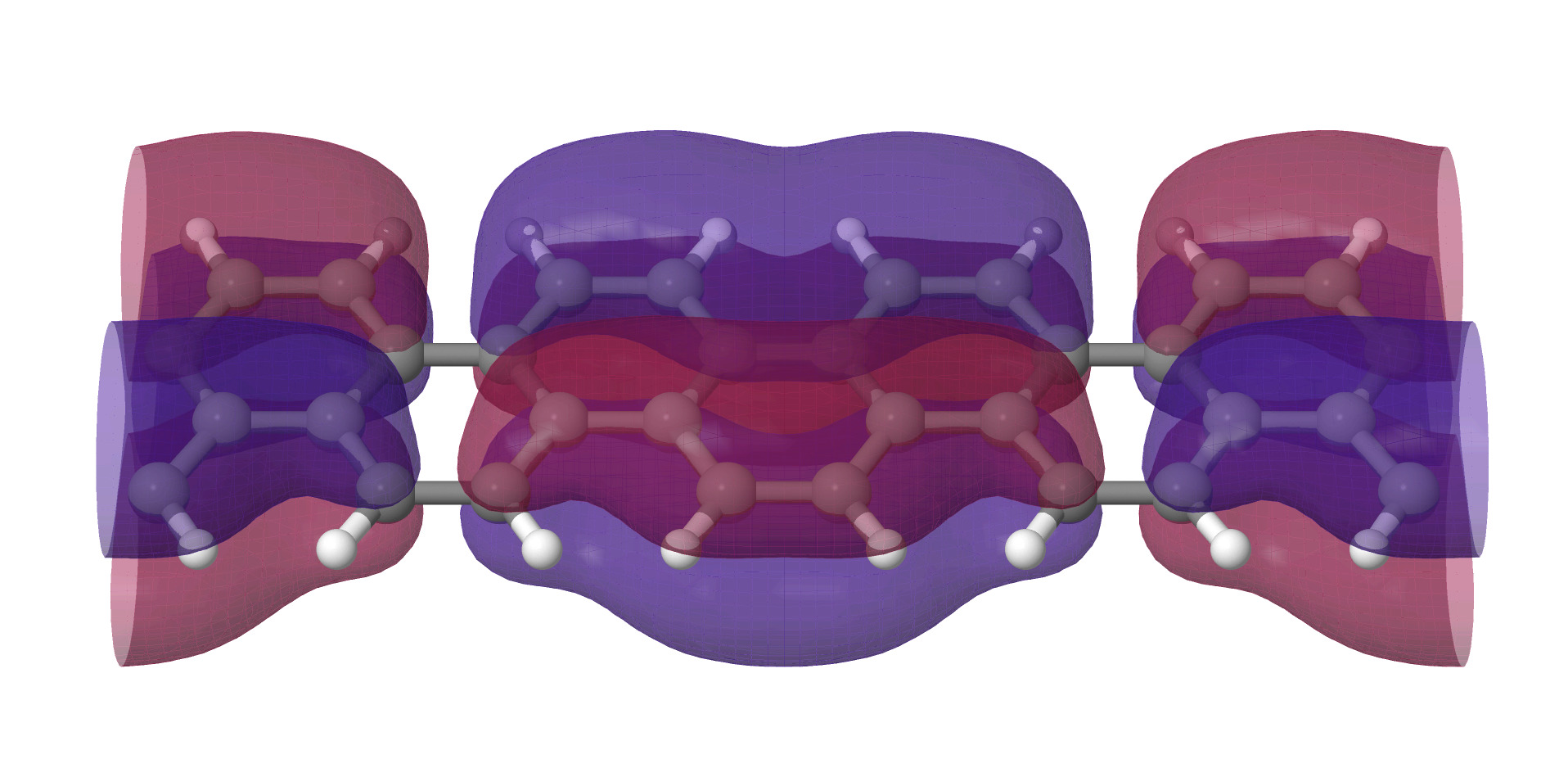} }

\subfloat[Foster--Boys Wannier\label{fig:nr-ML}]{\includegraphics[width=0.24\textheight]{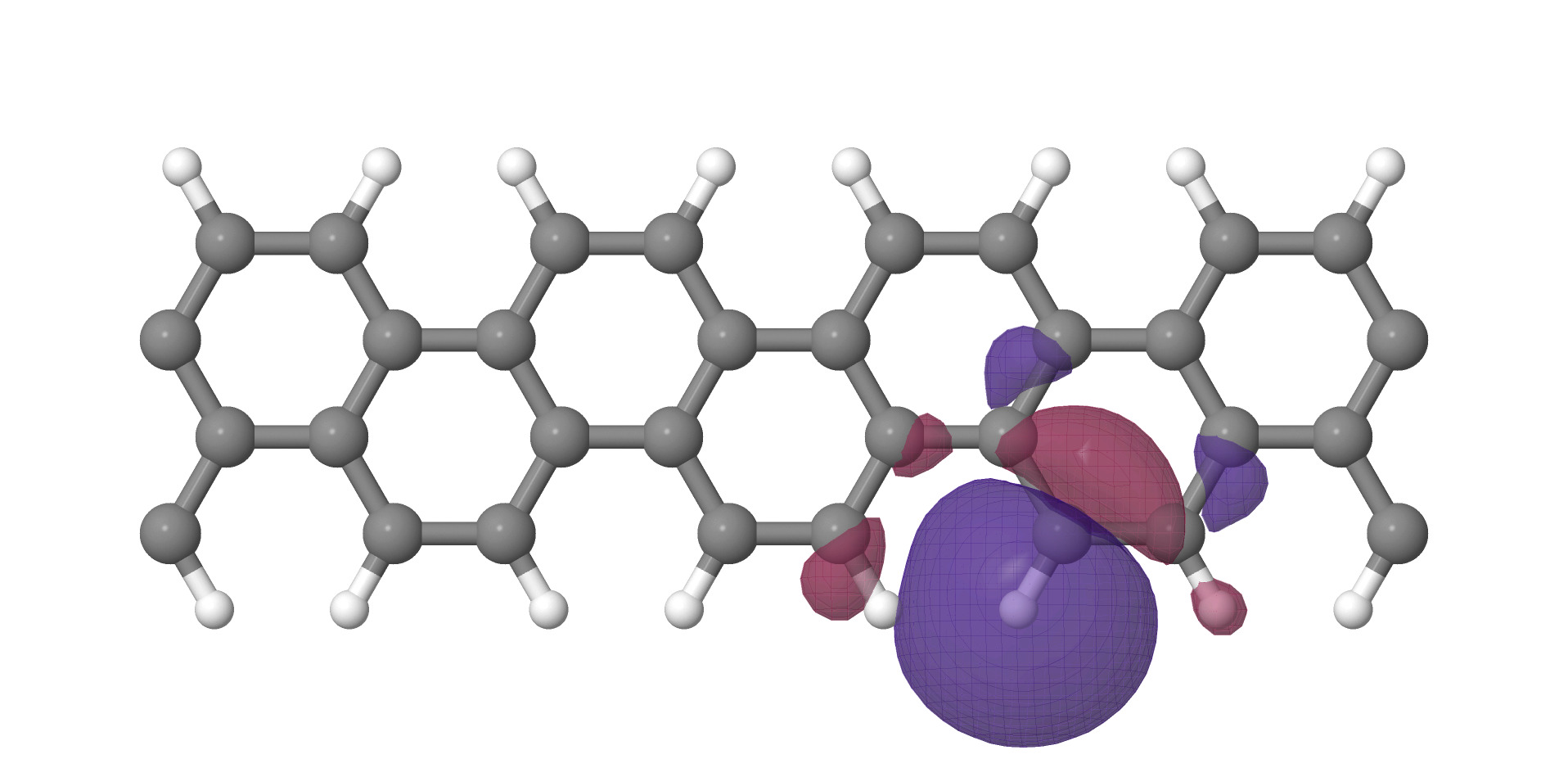}\includegraphics[width=0.24\textheight]{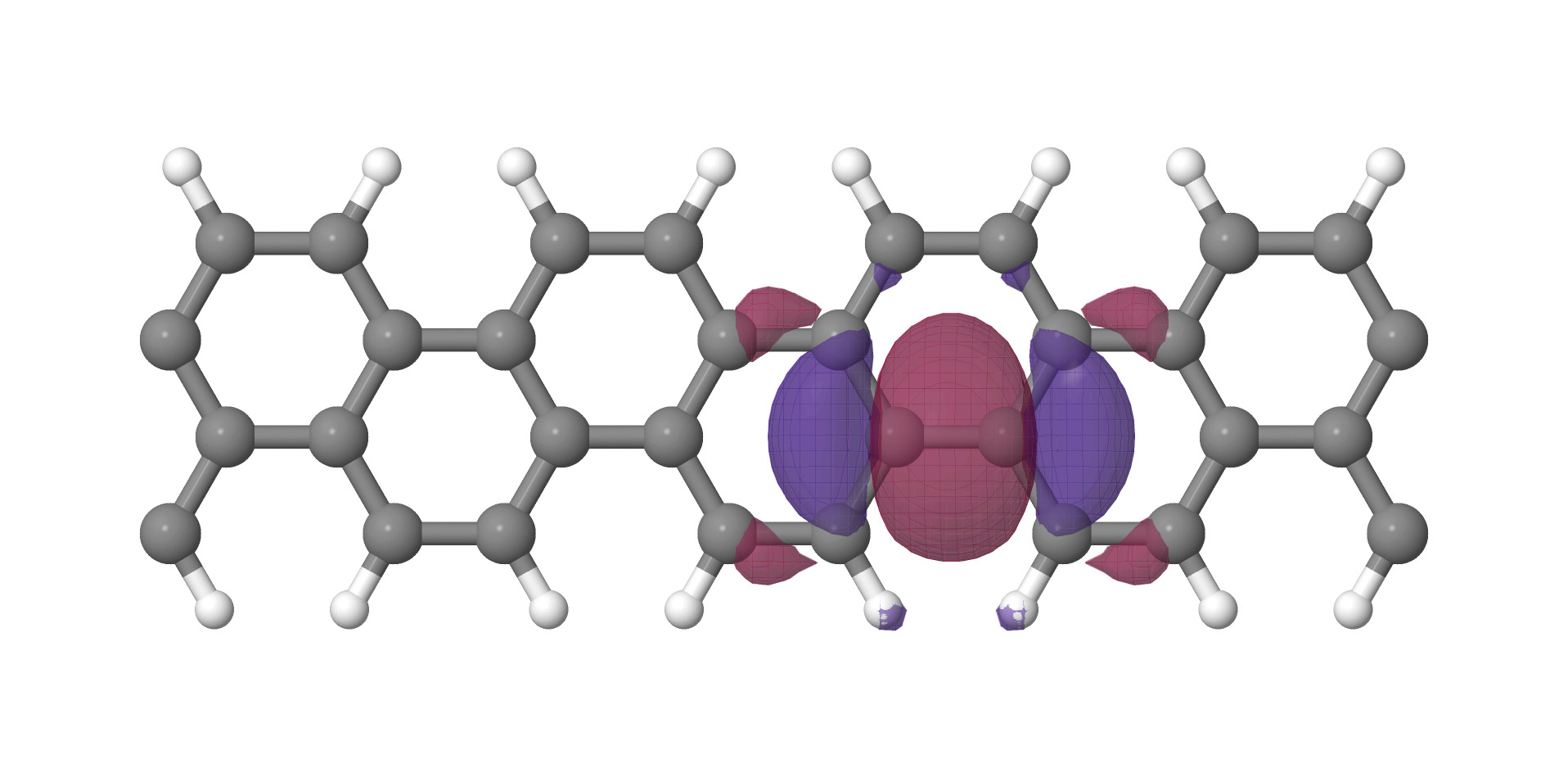}\includegraphics[width=0.24\textheight]{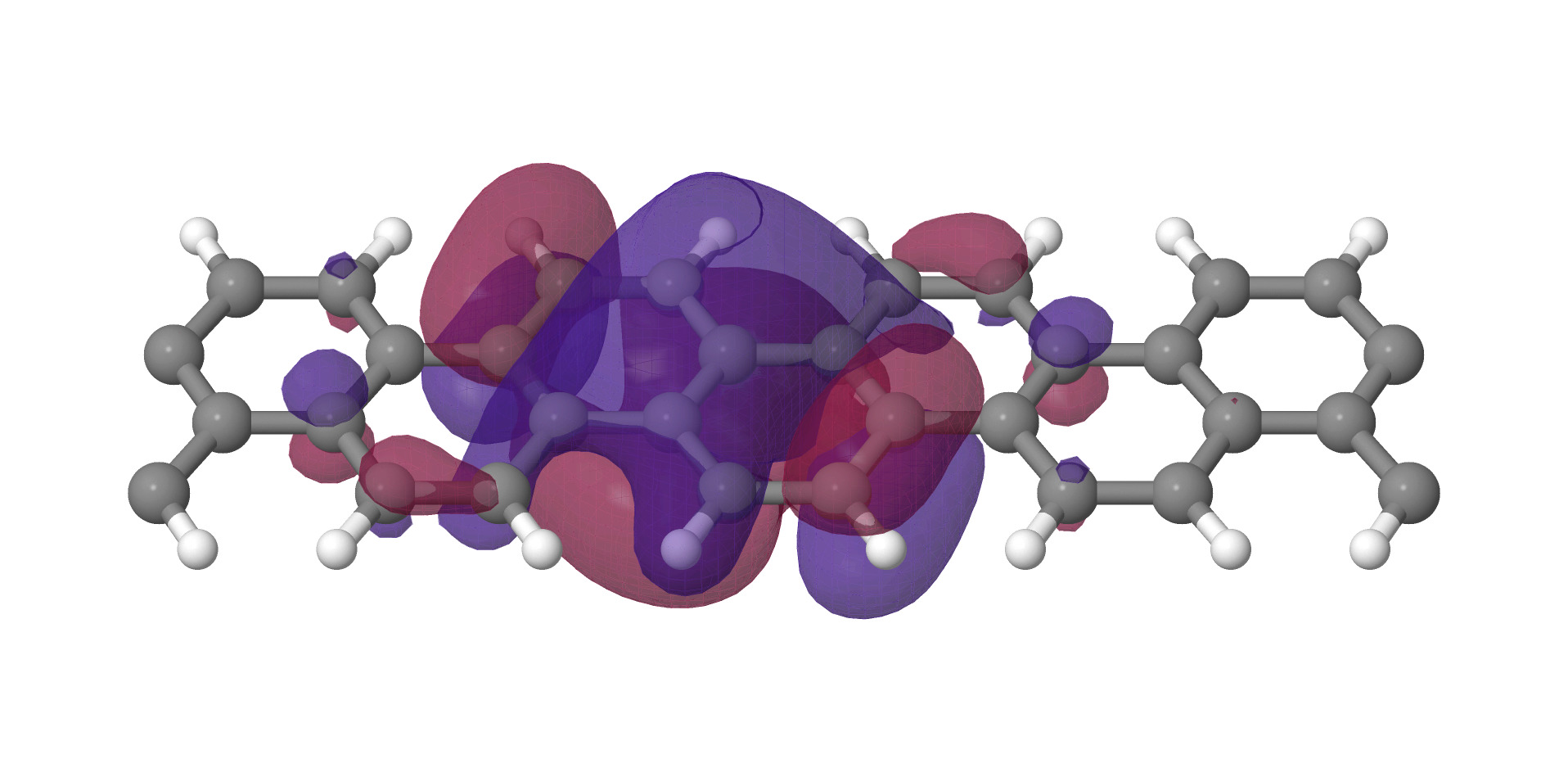}\includegraphics[width=0.24\textheight]{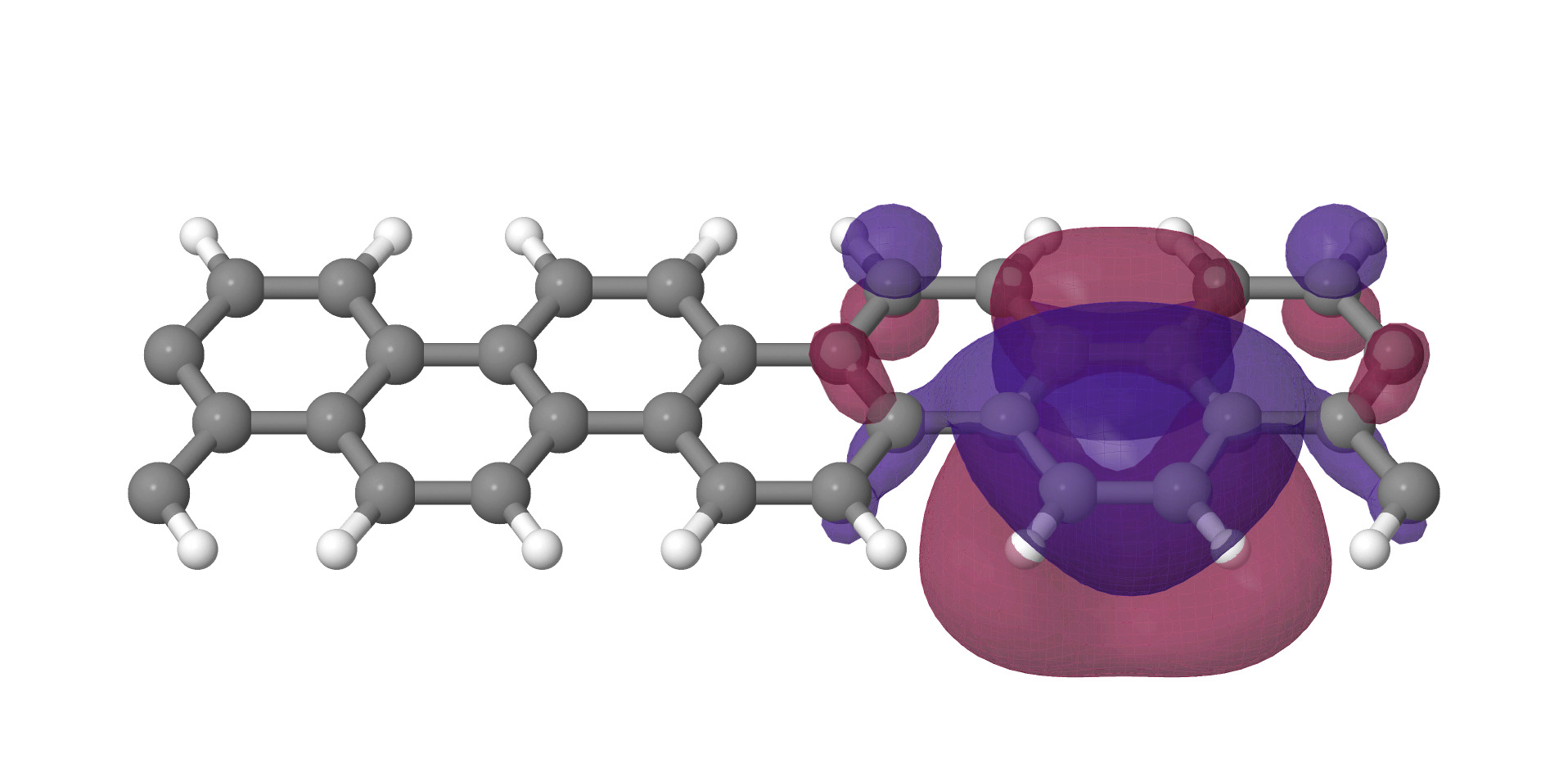} }

\subfloat[Pipek--Mezey Wannier\label{fig:nr-PM}]{\includegraphics[width=0.24\textheight]{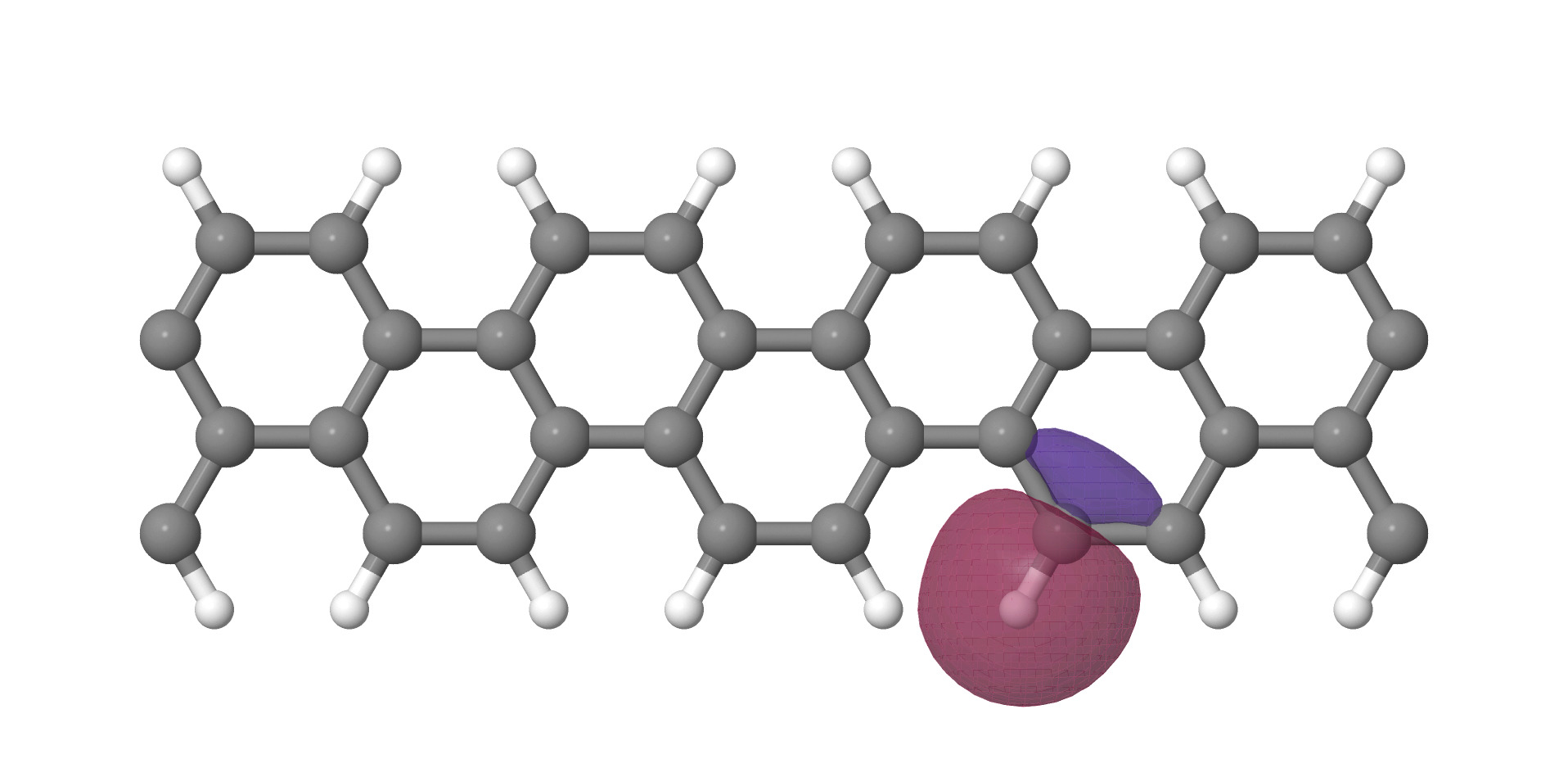}\includegraphics[width=0.24\textheight]{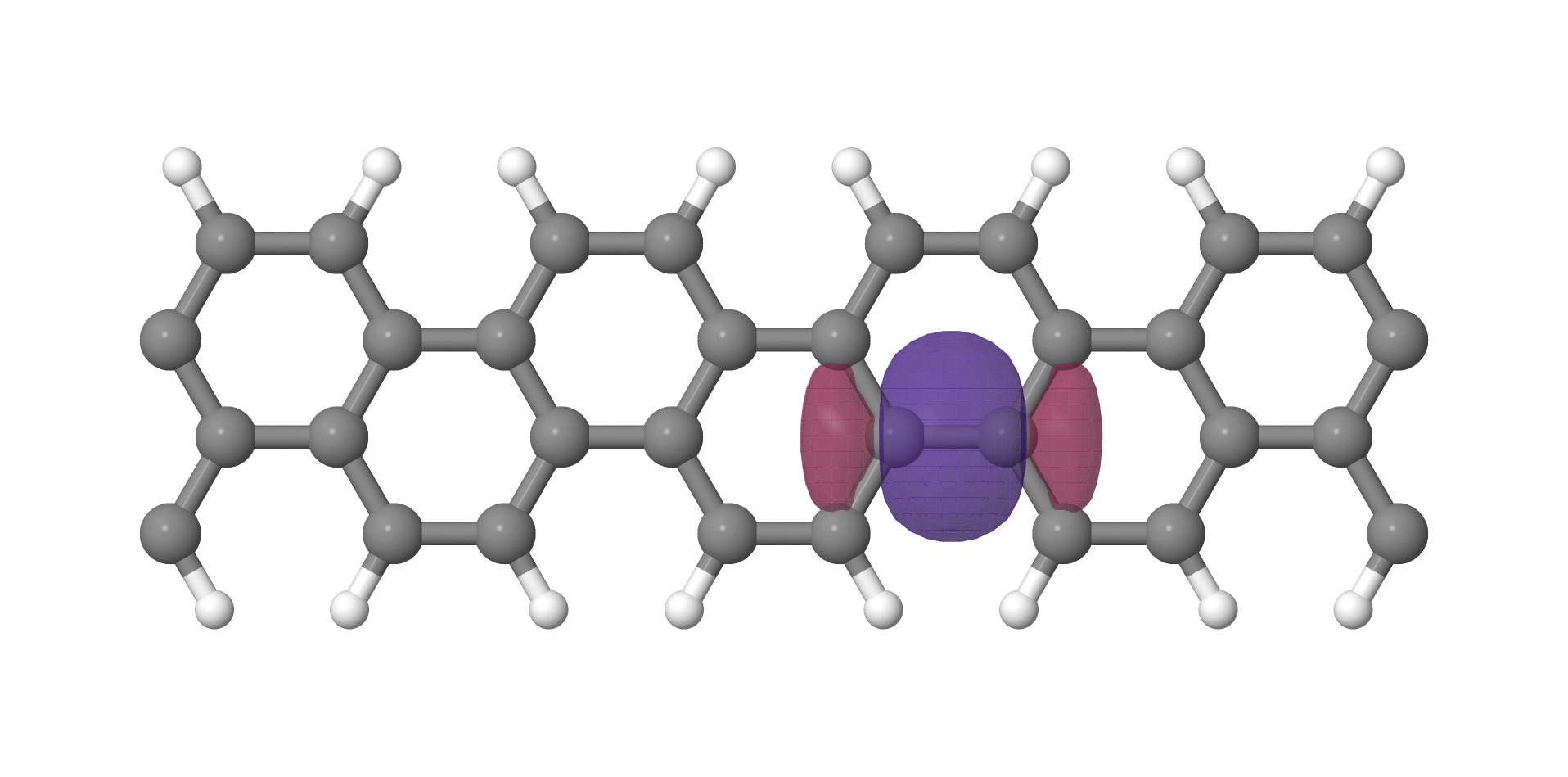}\includegraphics[width=0.24\textheight]{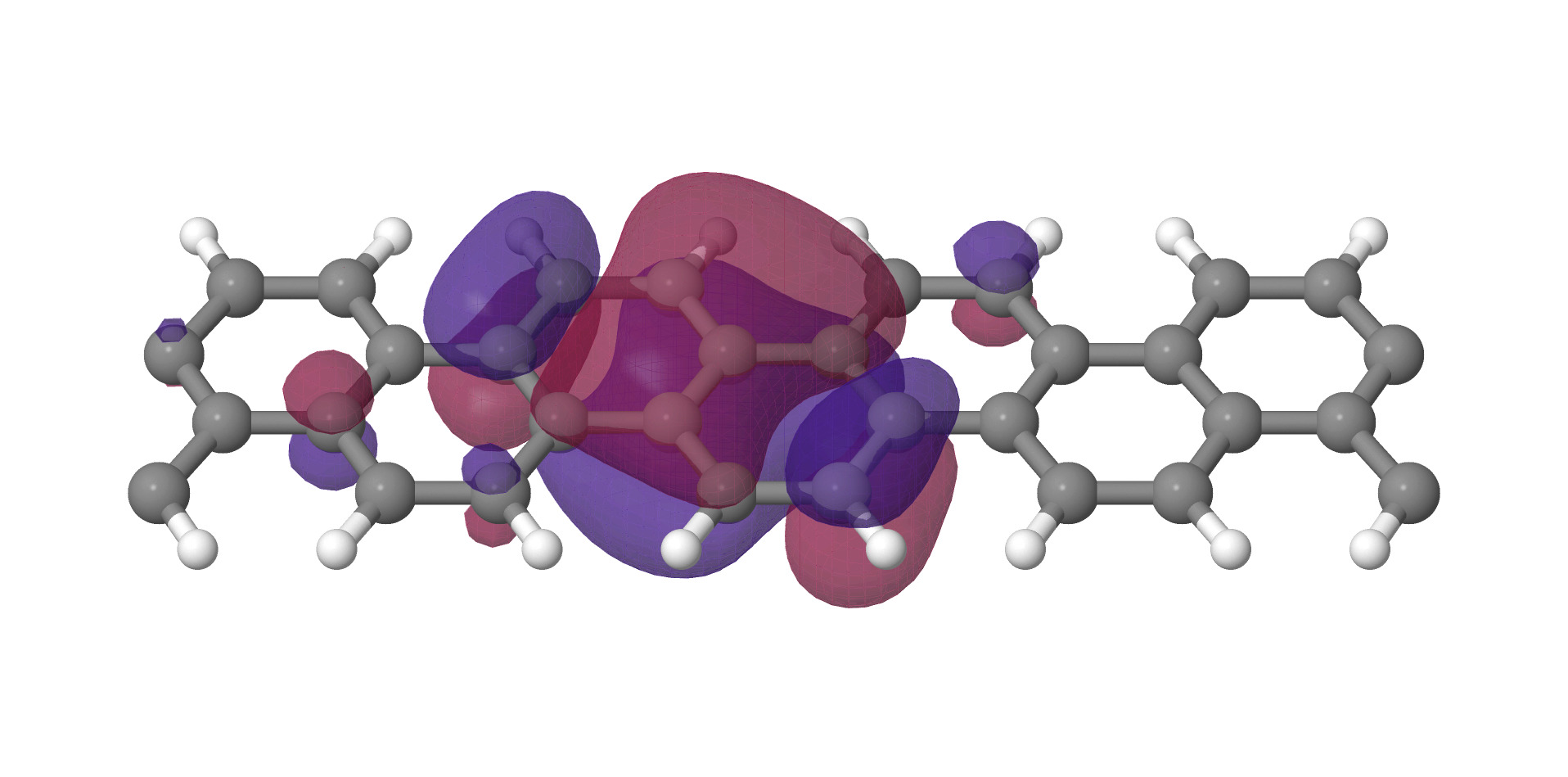}\includegraphics[width=0.24\textheight]{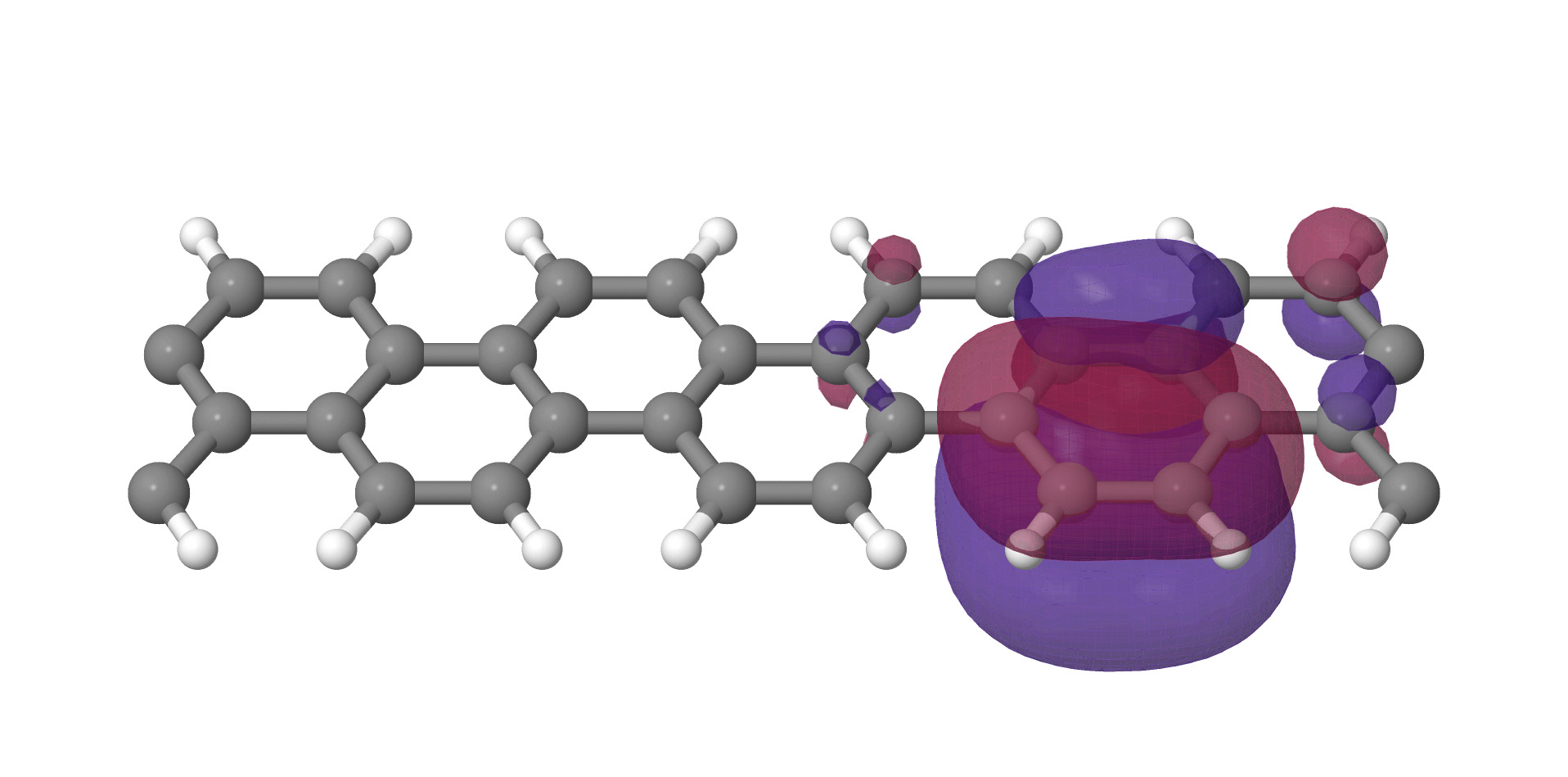} }

\caption{Kohn--Sham (top row), Foster--Boys Wannier function
  (middle row) and Pipek--Mezey Wannier function (bottom row) orbitals
  of (2,4)-armchair nanoribbon segment, subject to periodic boundary
  conditions (see main text).}
\label{fig:nanoribbon}

\end{sidewaysfigure*}

For the aromatic hydrocarbons -- benzene, coronene, supercoronene and
airmchair nanoribbons -- both the PMWF and FBWF methods reduce the set
of COs, which are distinct for each valence state, to a set of a few
highly localized orbitals representing the $\sigma$- and
$\pi$-bonds. \Figref{nanoribbon} presents example COs and LOs for a
(2,4)-armchair nanoribbon. The Kohn--Sham states spread over the
entire system. For the FBWF and PMWF methods, the first and second
column show example $\sigma$-bond orbitals, which are highly localized
carbon-hydrogen ($\sigma_\mr{CH}$)- and carbon-carbon
($\sigma_\mr{CC}$)-bonds, whereas the third and fourth columns show
examples of states with $\pi$ character, which are pure $\pi$ states for
PMWF but mixtures of $\sigma$- and $\pi$-bond orbitals in the case of
FBWF.

\begin{figure*}
\subfloat[Kohn--Sham\label{fig:cp-KS}]{\includegraphics[width=0.33\textwidth]{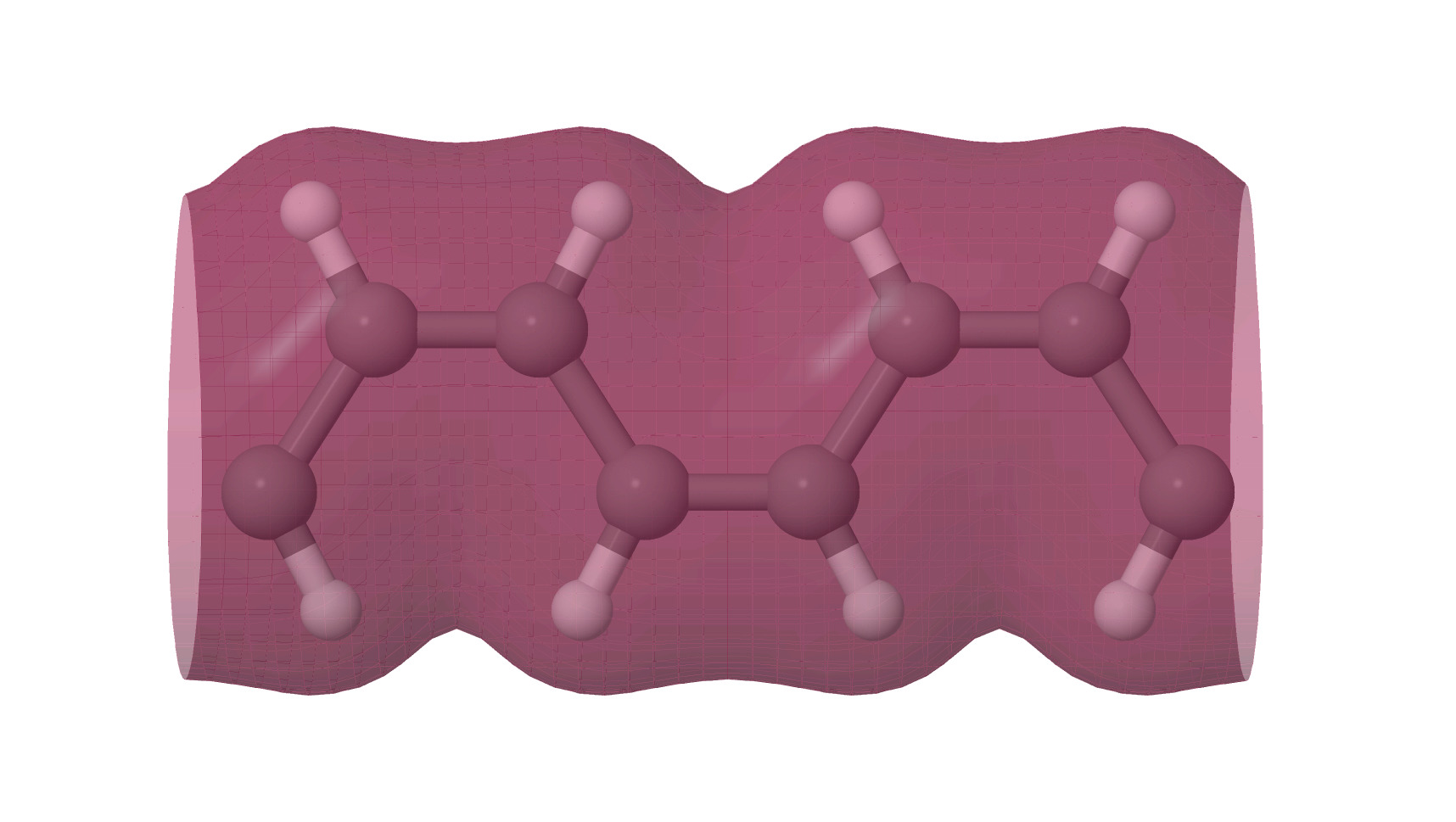}\includegraphics[width=0.33\textwidth]{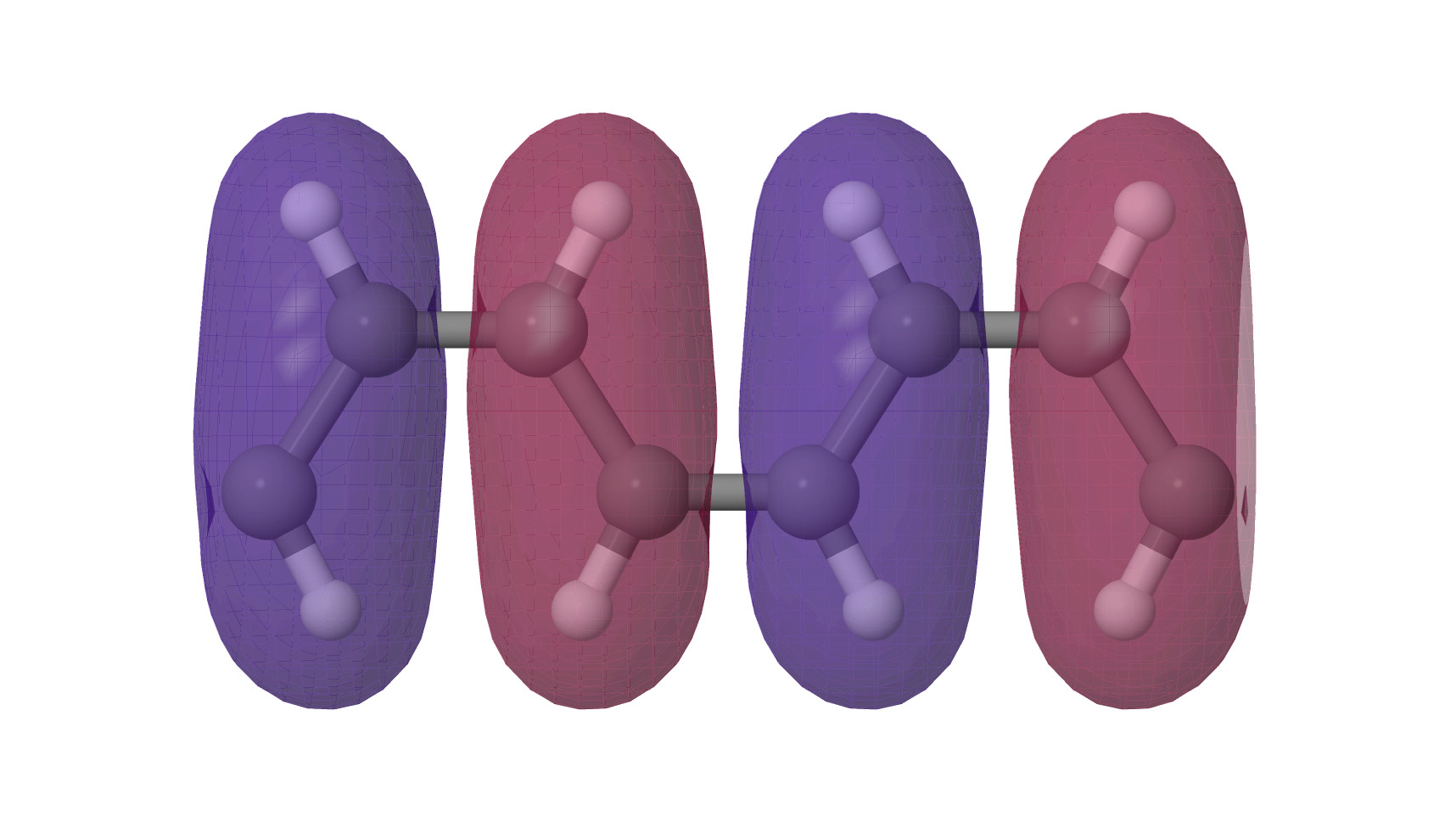}\includegraphics[width=0.33\textwidth]{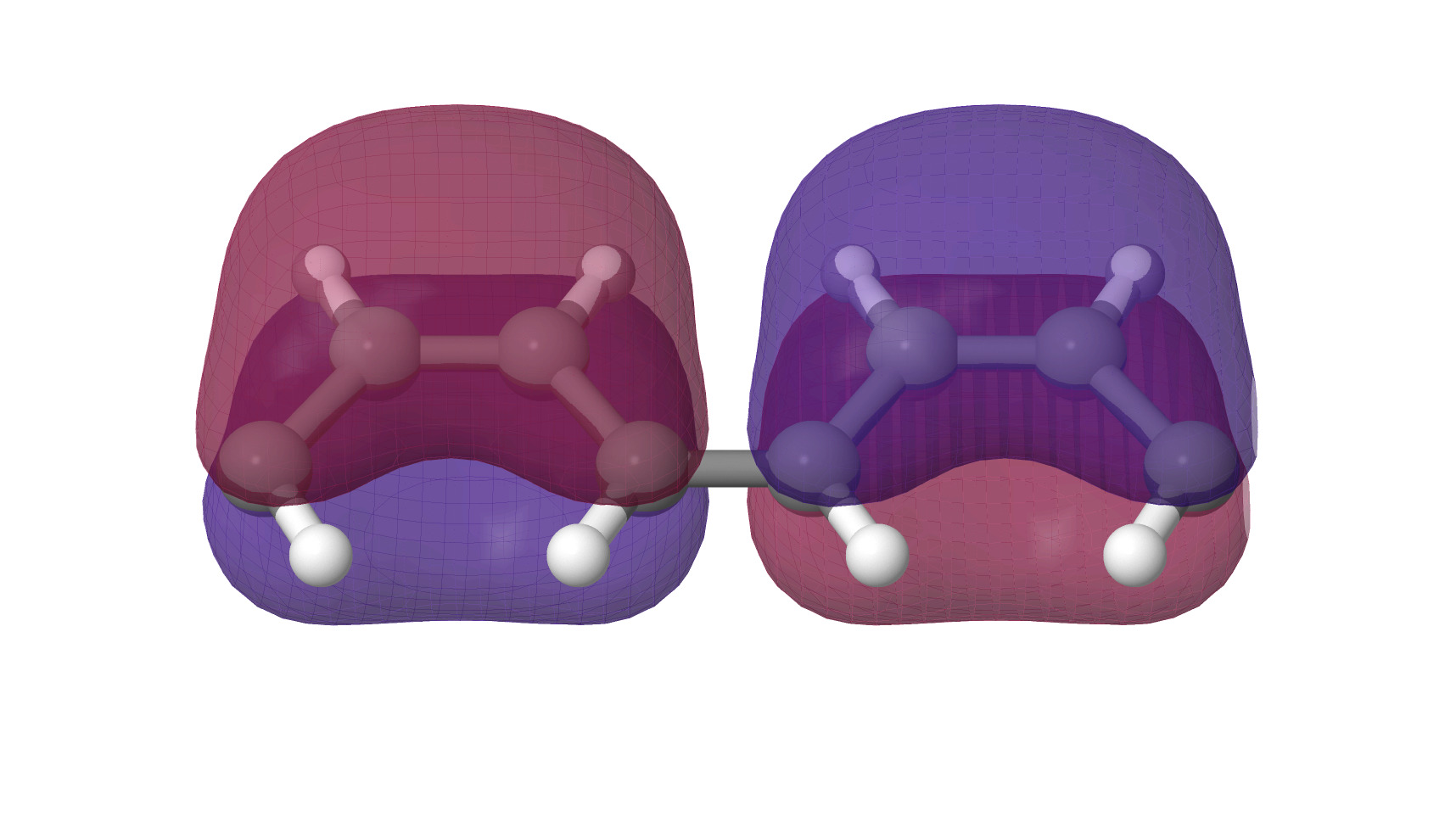} }

\subfloat[Foster--Boys Wannier\label{fig:cp-ML}]{\includegraphics[width=0.33\textwidth]{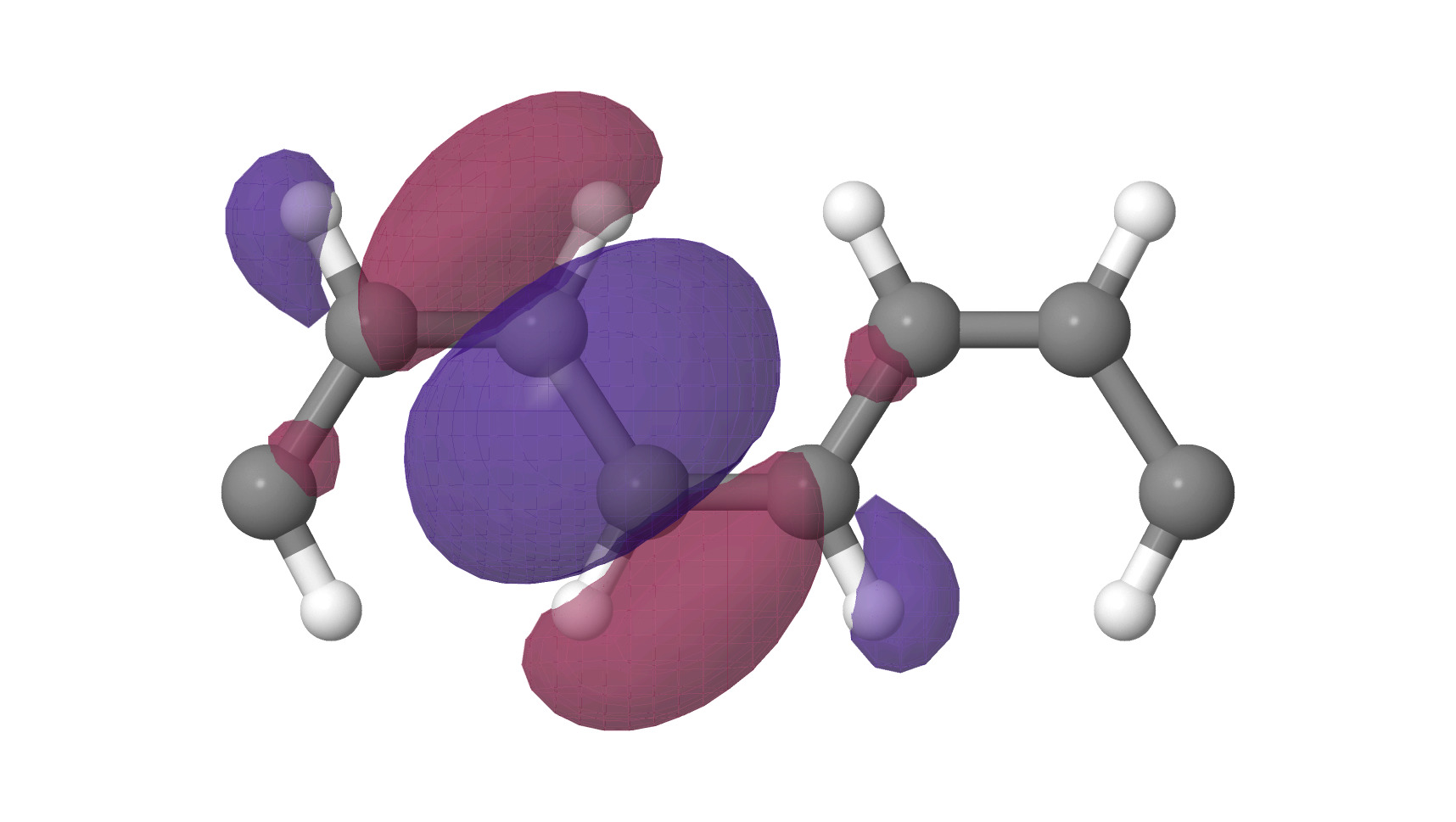}\includegraphics[width=0.33\textwidth]{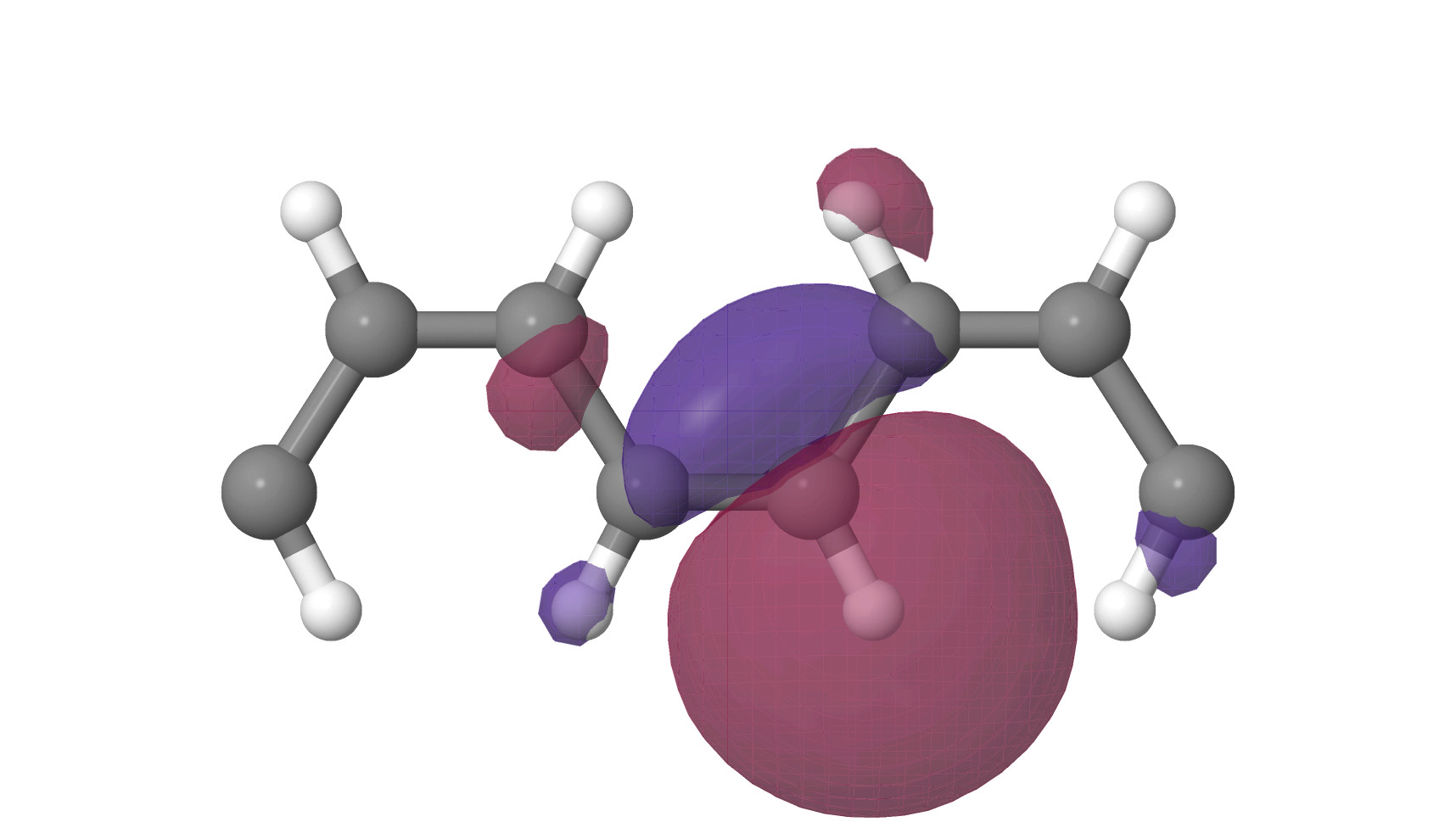}\includegraphics[width=0.33\textwidth]{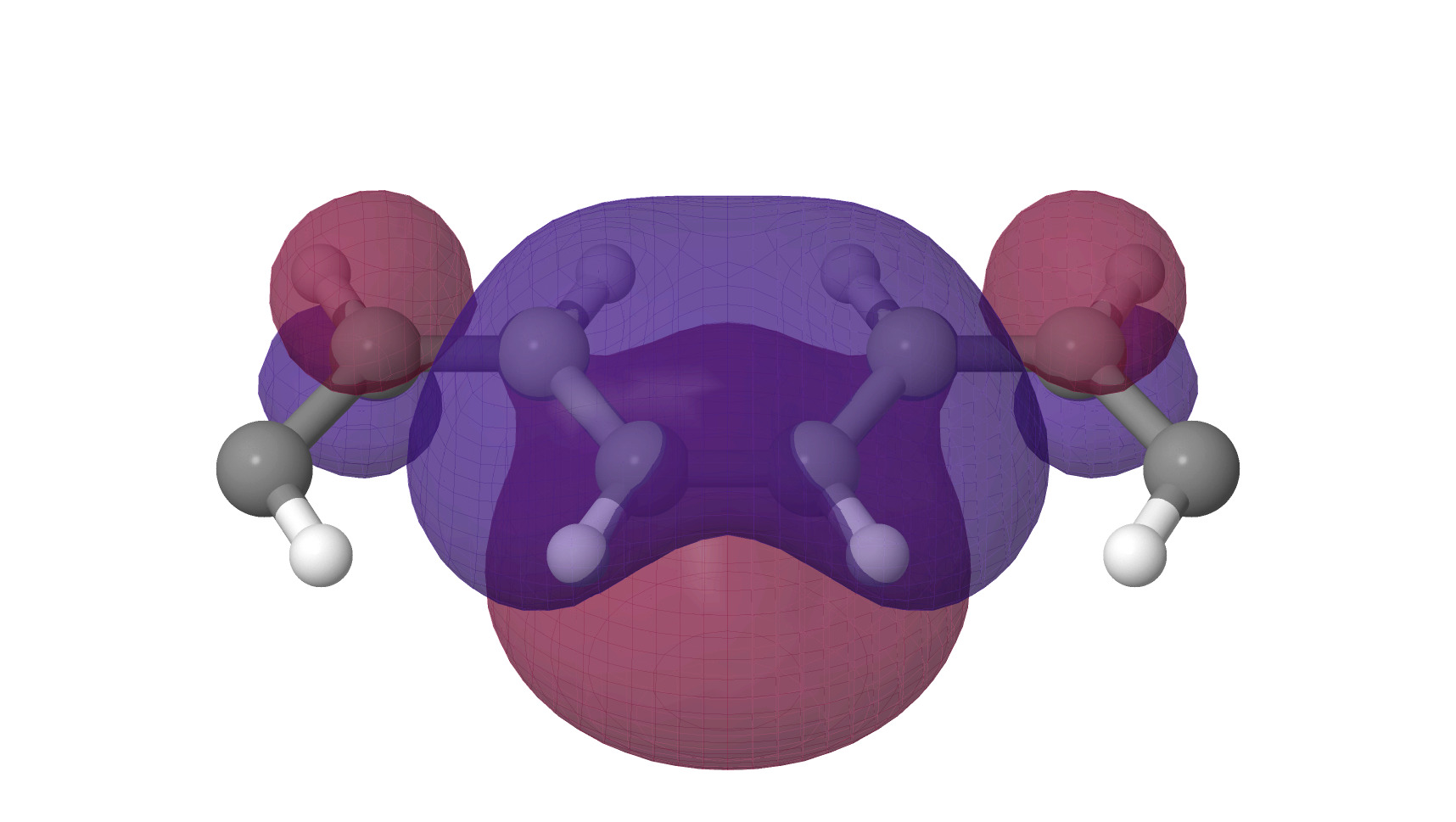} }

\subfloat[Pipek--Mezey Wannier\label{fig:cp-PM}]{\includegraphics[width=0.33\textwidth]{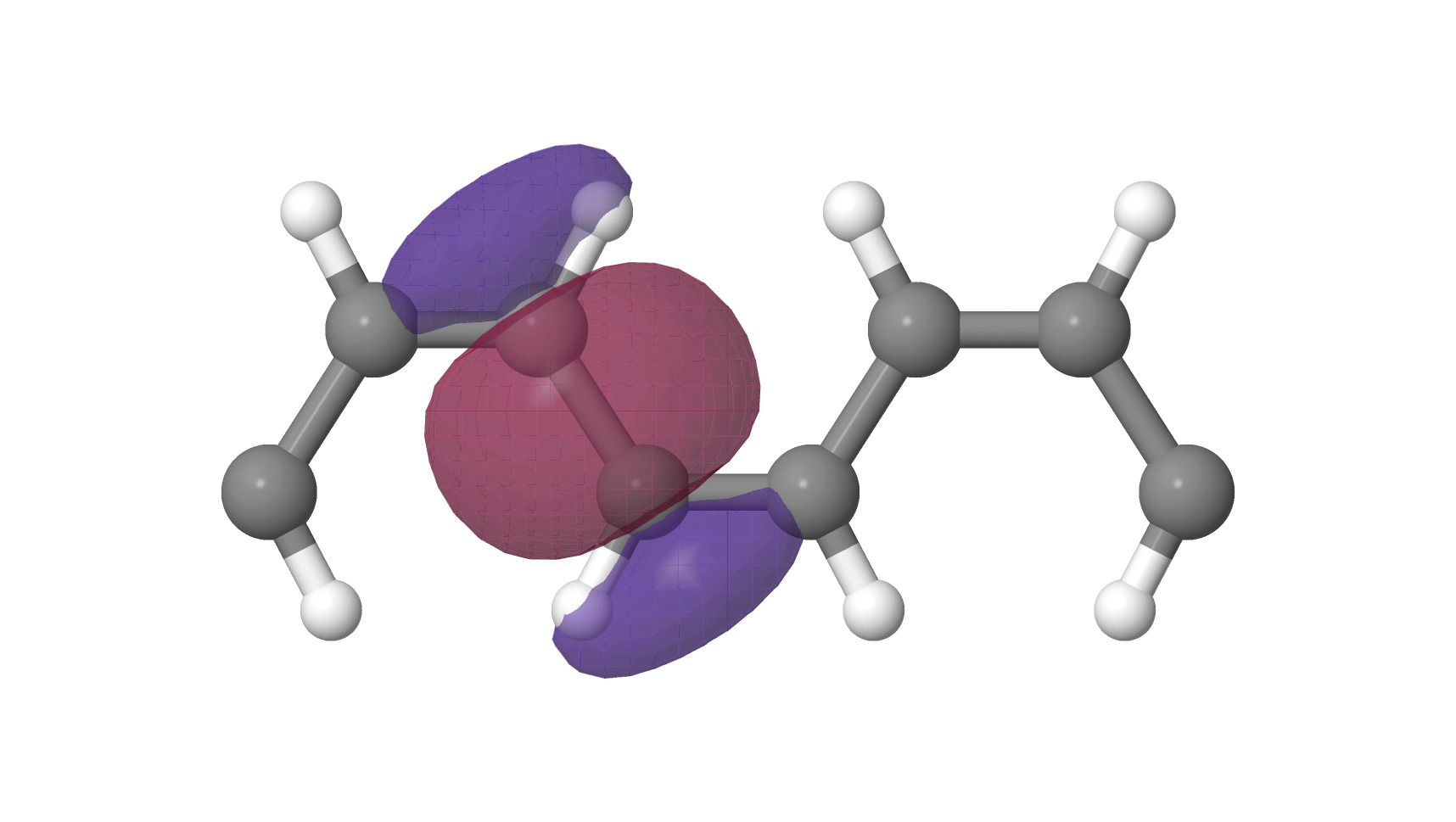}\includegraphics[width=0.33\textwidth]{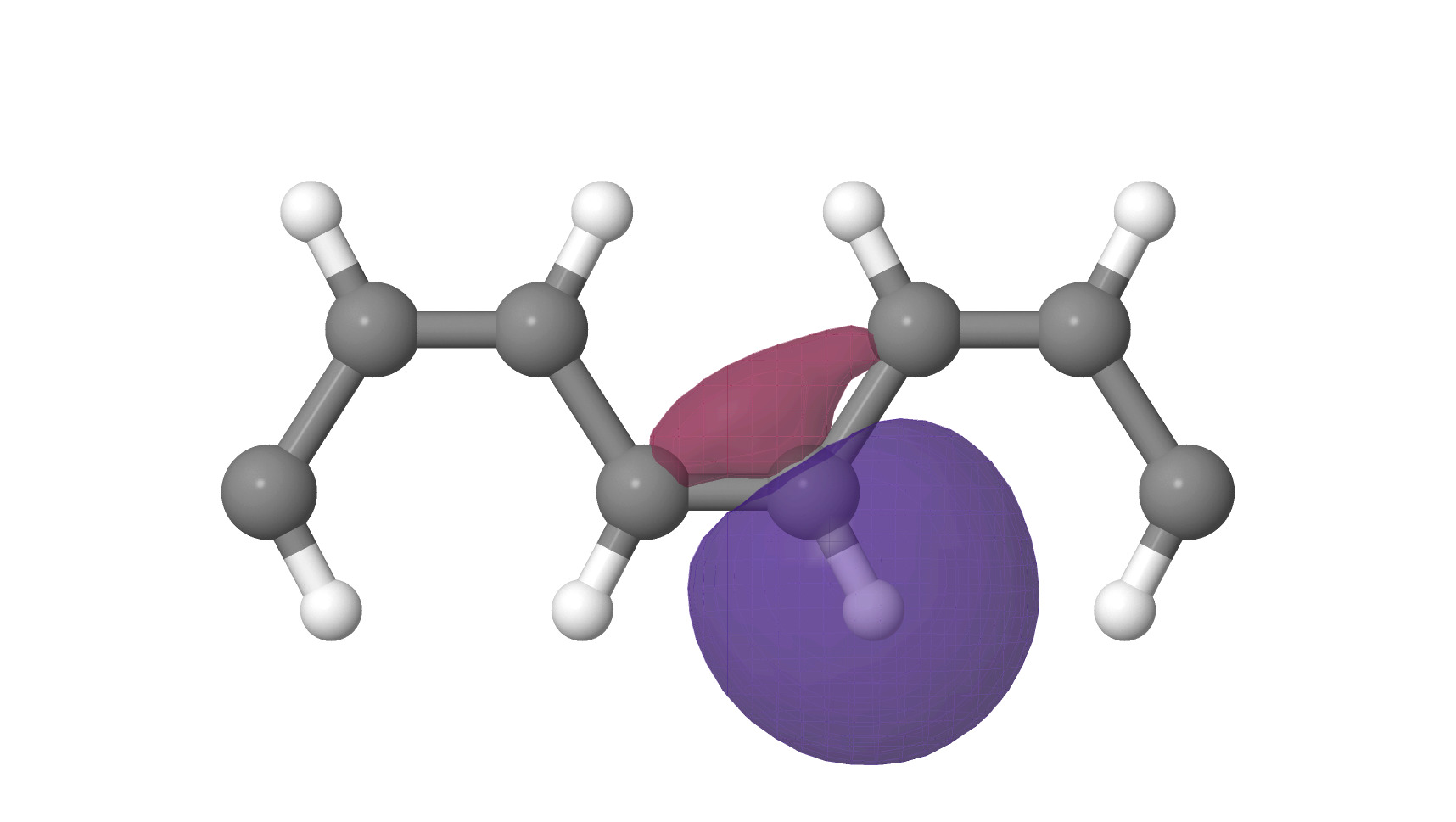}\includegraphics[width=0.33\textwidth]{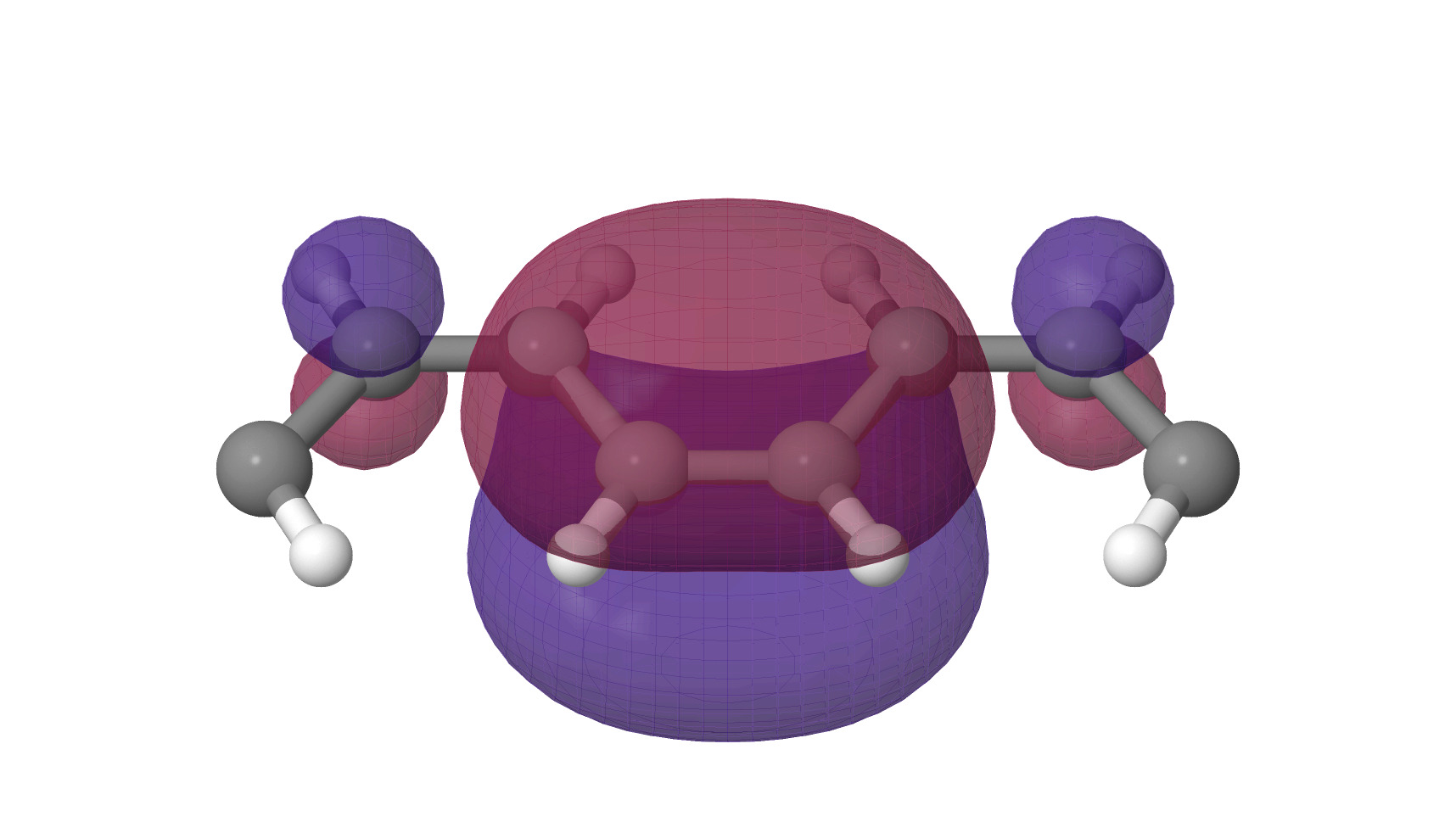} }

  \caption{Kohn--Sham (top row), Foster--Boys Wannier function
    (middle row) and Pipek--Mezey Wannier function (bottom row)
    orbitals of cis-polyacetylene, represented by a
    -[C$_8$H$_8$]- segment, subject to periodic boundary
    conditions (see main text).  }
  \label{fig:cispoly}
\end{figure*}

COs, FBWFs and PMWFs for cis-polyacetylene are presented in
\figref{cispoly}. The FBWF and PMWF methods give similar orbitals
corresponding to $\sigma_\mr{CC}$ (first column) and $\sigma_\mr{CH}$
bonds (middle column), but the FBWFs do not appear as localized
as the PMWFs as judged by the 75\% density isosurfaces. Only four
$\sigma_\mr{CC}$ bonds of the eight possible $\sigma_\mr{CC}$-bonds in
the system are obtained with the FBWF method, while the other four are
mixed with the $\pi_\mr{CC}$-bonds to form $\tau$ states as shown in
the third column. The PMWF set consists of eight $\sigma_\mr{CC}$-,
eight $\sigma_\mr{CH}$- and four $\pi_\mr{CC}$-bonds, in accordance
with the chemical picture of such a segment of cis-polyacetylene.

\begin{figure*}
\subfloat[Kohn--Sham\label{fig:ca-KS}]{\includegraphics[width=0.33\textwidth]{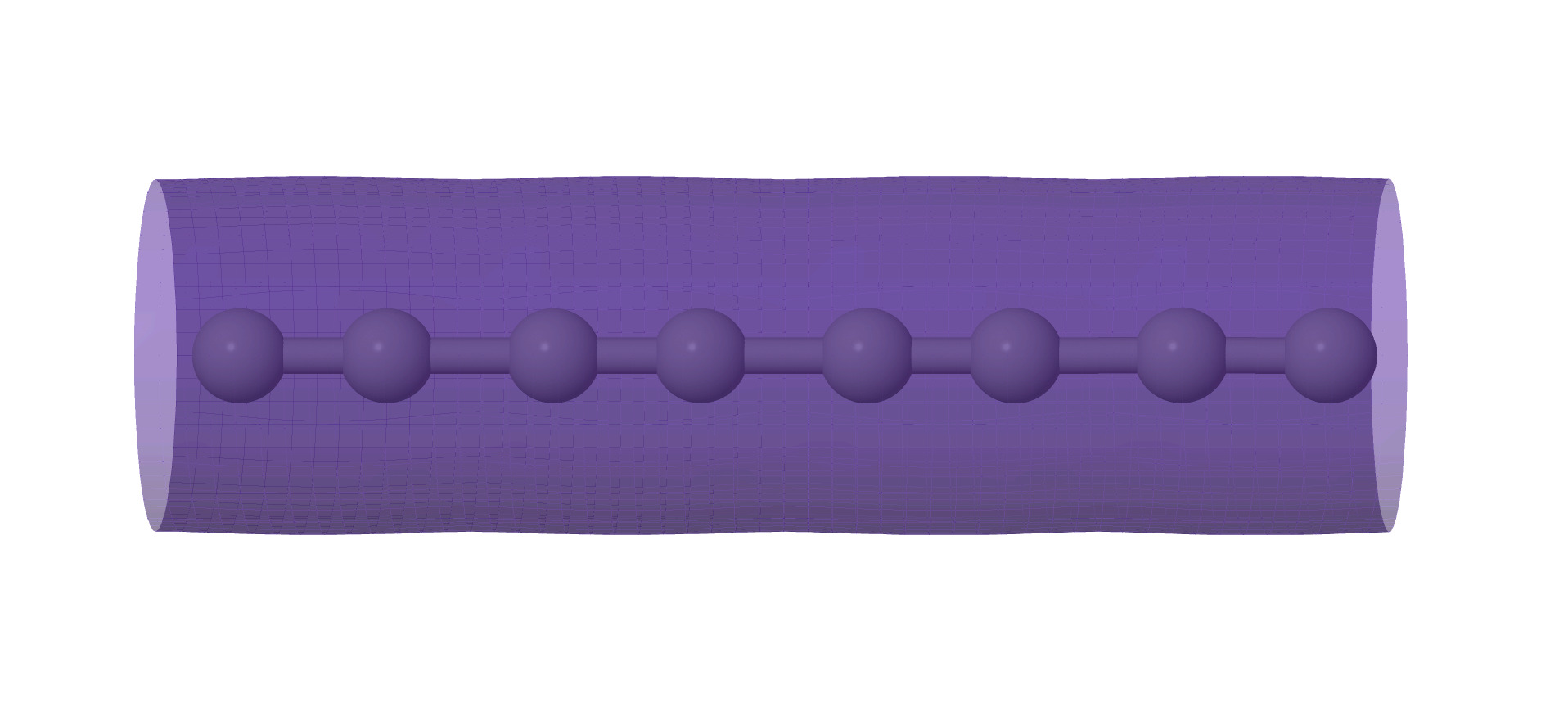}\includegraphics[width=0.33\textwidth]{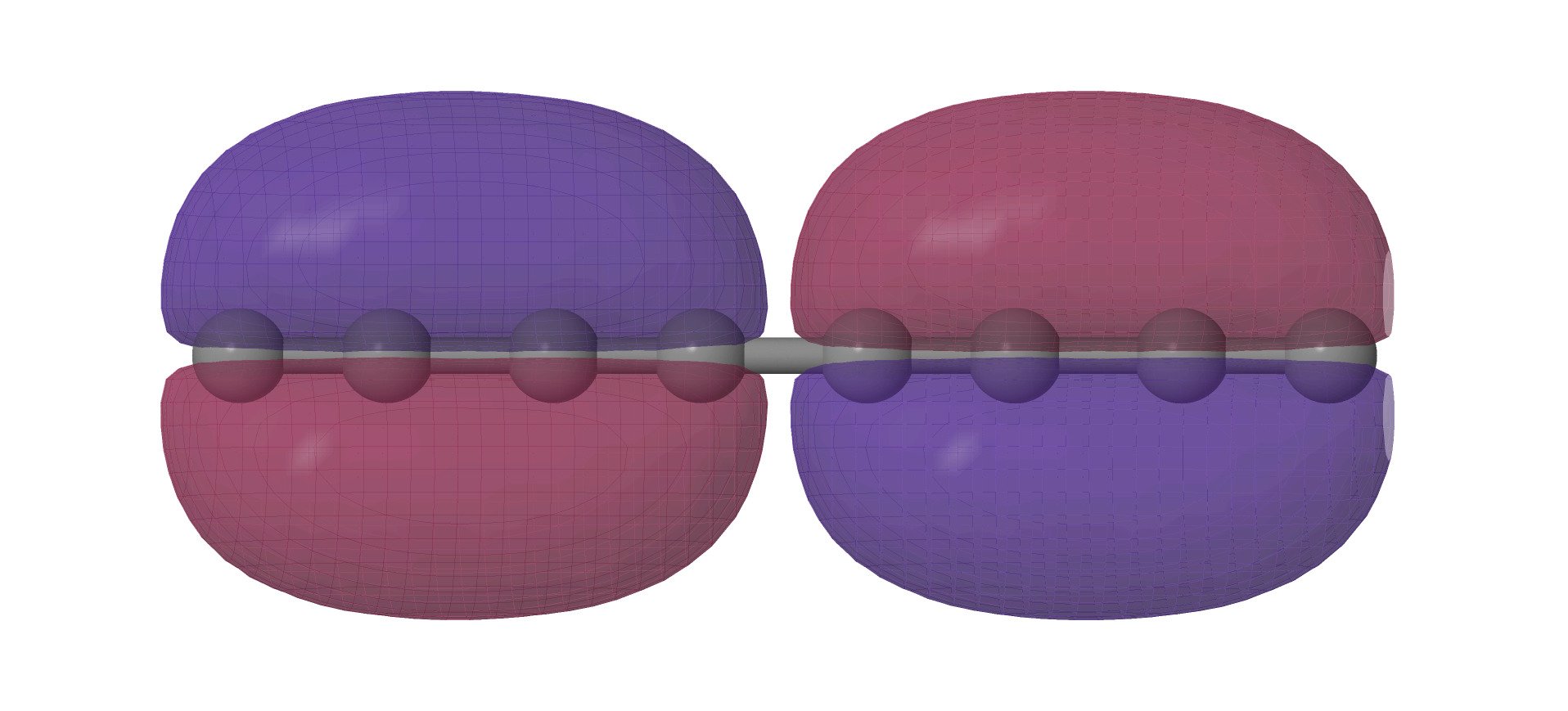} }

\subfloat[Foster--Boys Wannier\label{fig:ca-ML}]{\includegraphics[width=0.33\textwidth]{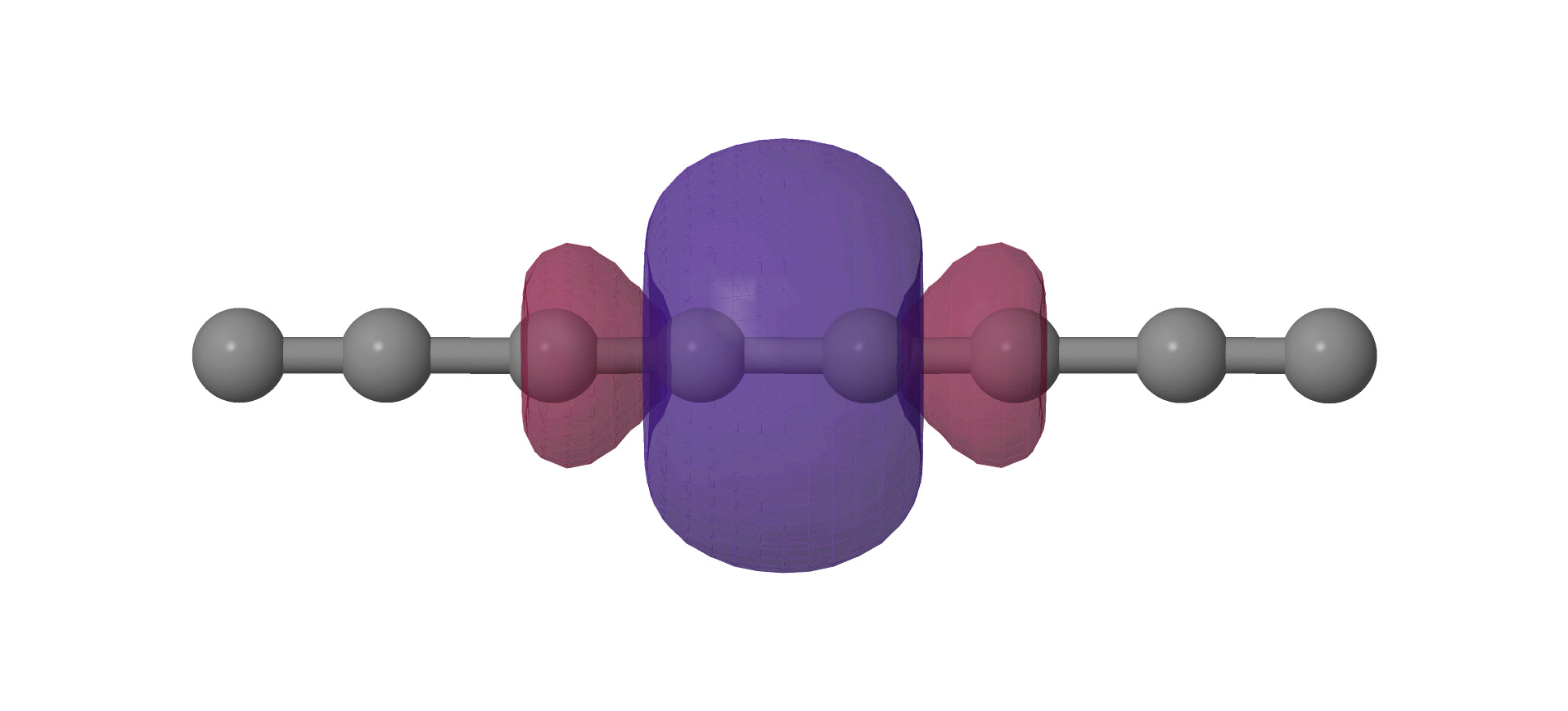}\includegraphics[width=0.33\textwidth]{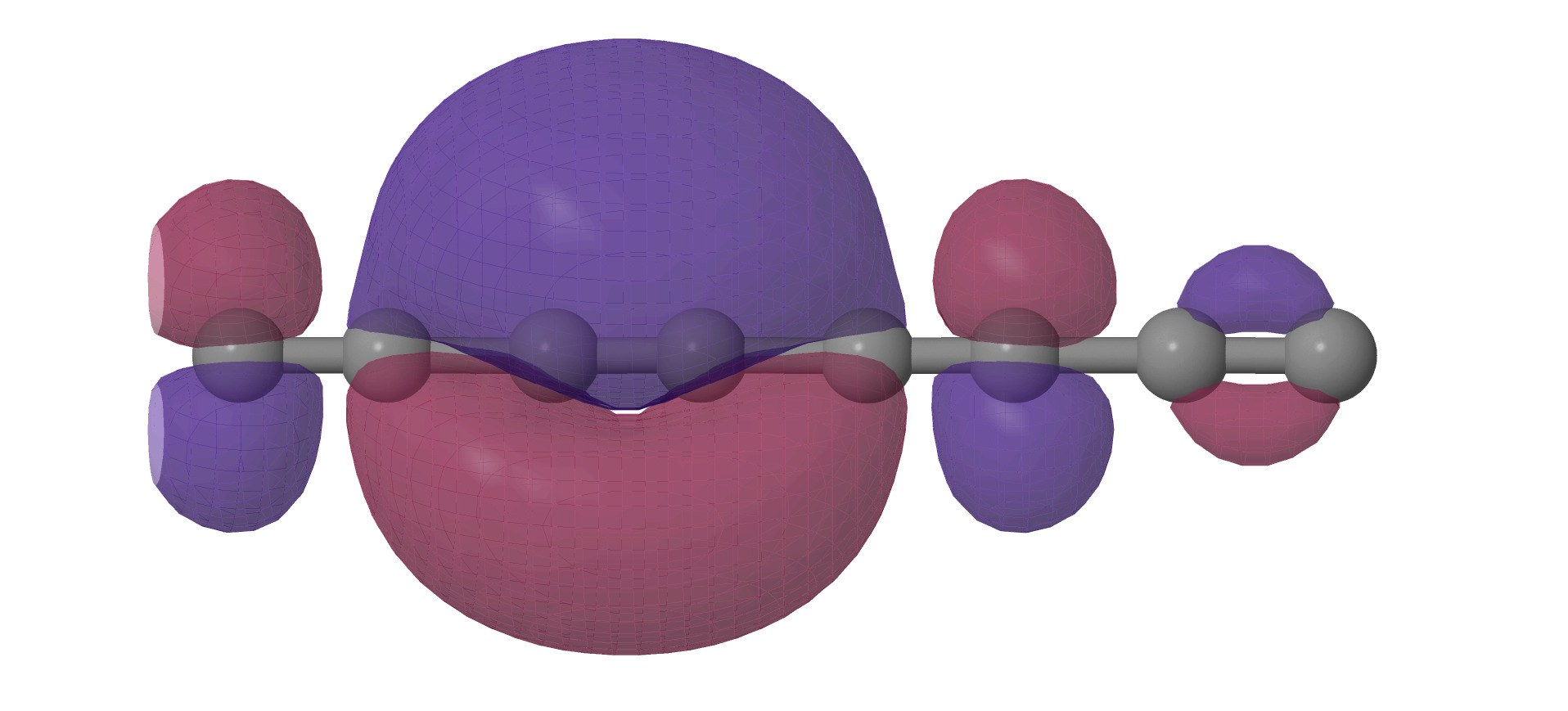} }

\subfloat[Pipek--Mezey Wannier\label{fig:ca-PM}]{\includegraphics[width=0.33\textwidth]{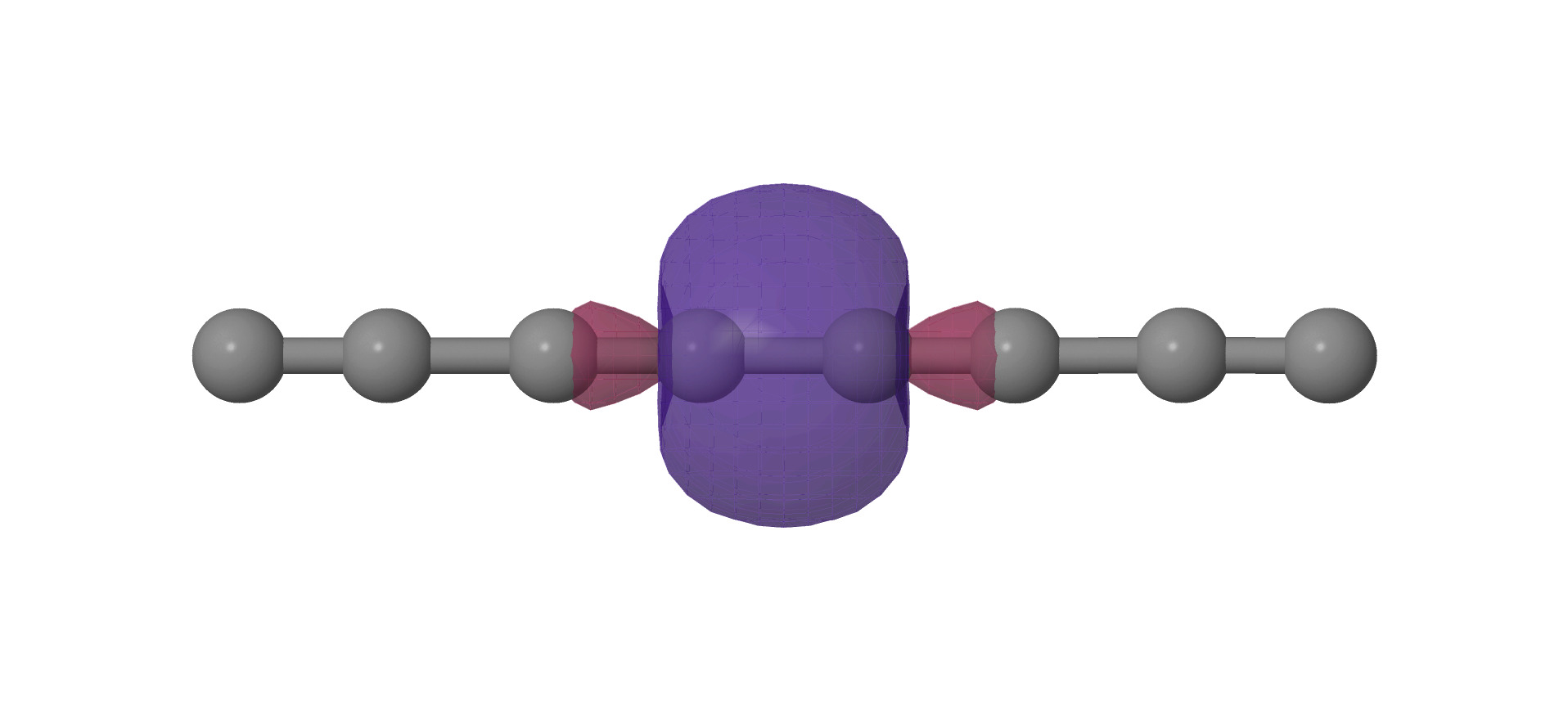}\includegraphics[width=0.33\textwidth]{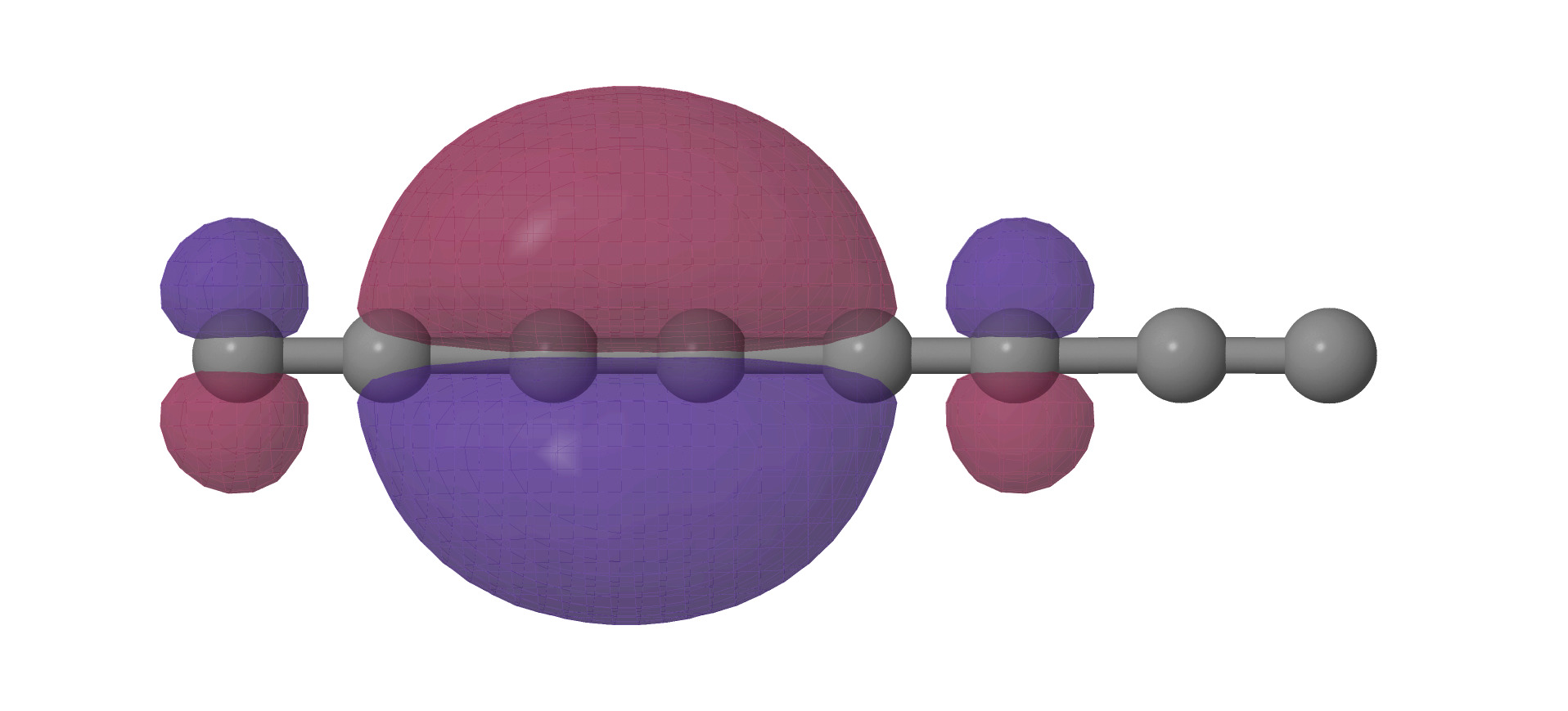} }

  \caption{Kohn--Sham (top row), Foster--Boys Wannier function
    (middle row) and Pipek--Mezey Wannier function (bottom row)
    orbitals of carbyne, represented by a
    -[C$_8$]- segment subject to periodic boundary
    conditions (see main text).  }
  \label{fig:carbyne}
\end{figure*}

\Figref{carbyne} presents COs, FBWFs and PMWFs for carbyne. Here,
there are in total 16 COs, half of them of $\sigma$ character and the
other half of $\pi$ character. The FBWFs reduce to a set of two
distinct types of orbitals, one representing the $\sigma_\mr{CC}$-bond
(first column) and the other being a set of $\tau$ mixed states
(second column) that are a 33/66 mixture of $\sigma$ and $\pi$
COs. This composition arises because there are two $\pi$-bonds for
every two carbon atoms, as carbyne has alternating single and triple
bonds.  Due to this mixing, only four of eight possible
$\sigma_\mr{CC}$ bonds are represented by the set of FBWFs, with the
other four being mixed with the $\pi$-bond orbitals to form twelve
mixed $\tau$-bond orbitals. In contrast, the set of PMWFs consists of
eight $\sigma_\mr{CC}$ and eight $\pi_\mr{CC}$ bond orbitals,
describing all possible carbon-carbon single and alternating triple
bonds, matching the chemical picture of the carbyne segment.

\begin{figure*}
\subfloat[Foster--Boys Wannier\label{fig:bc-ML}]{\includegraphics[width=0.33\textwidth]{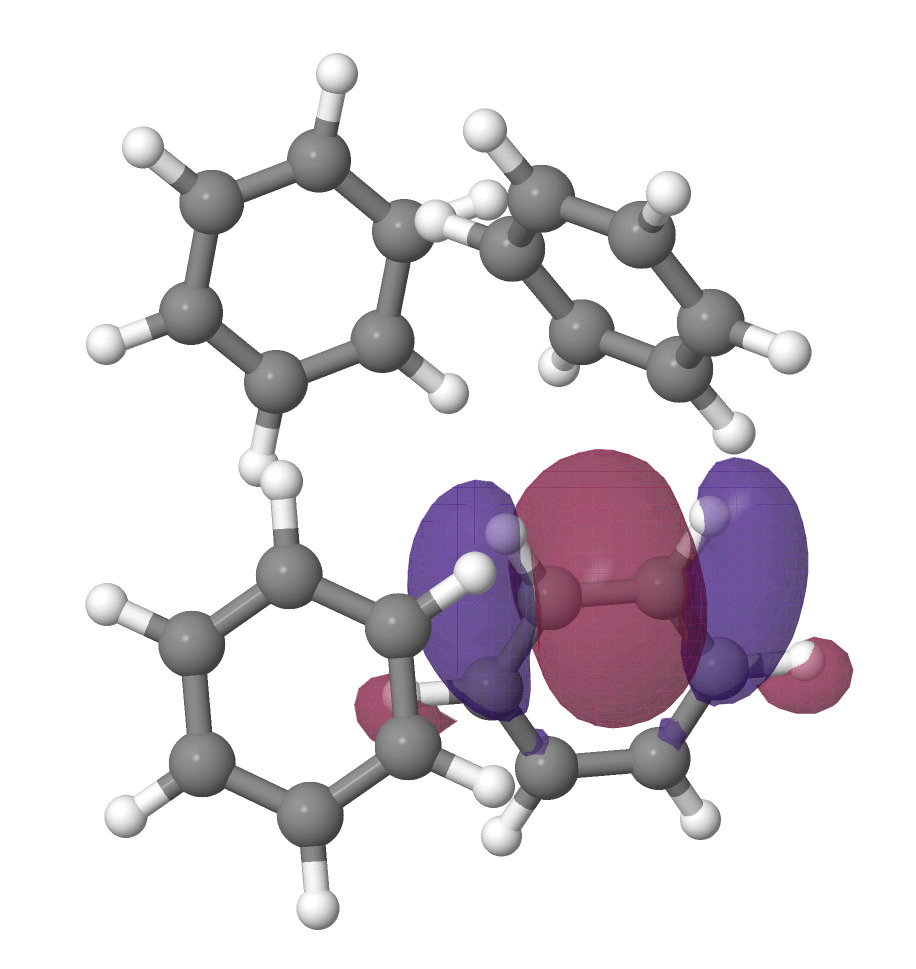}\includegraphics[width=0.33\textwidth]{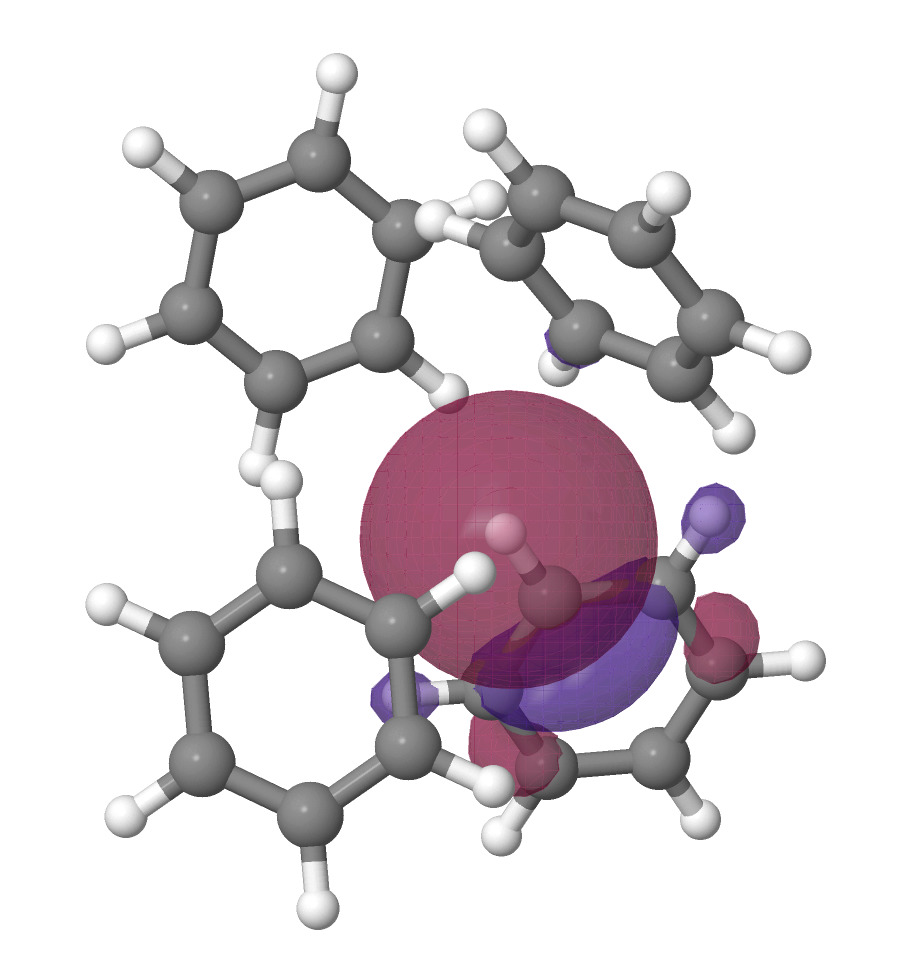}\includegraphics[width=0.33\textwidth]{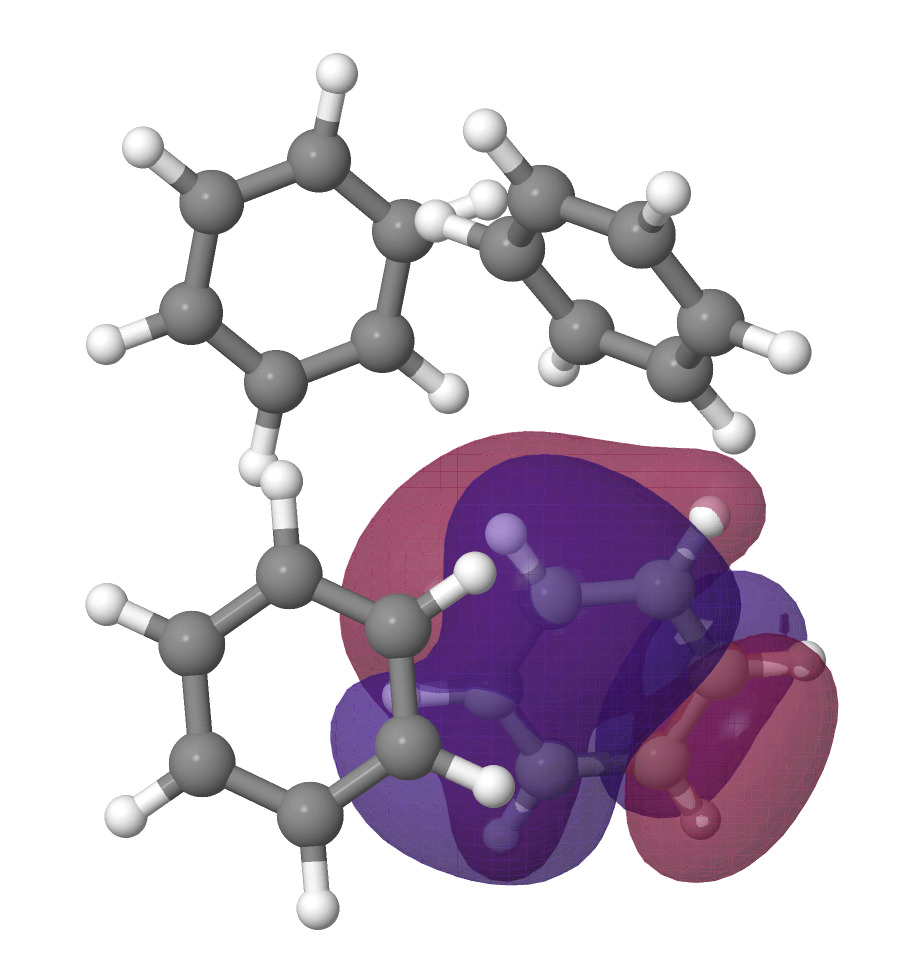} }

\subfloat[Pipek--Mezey Wannier\label{fig:bc-PM}]{\includegraphics[width=0.33\textwidth]{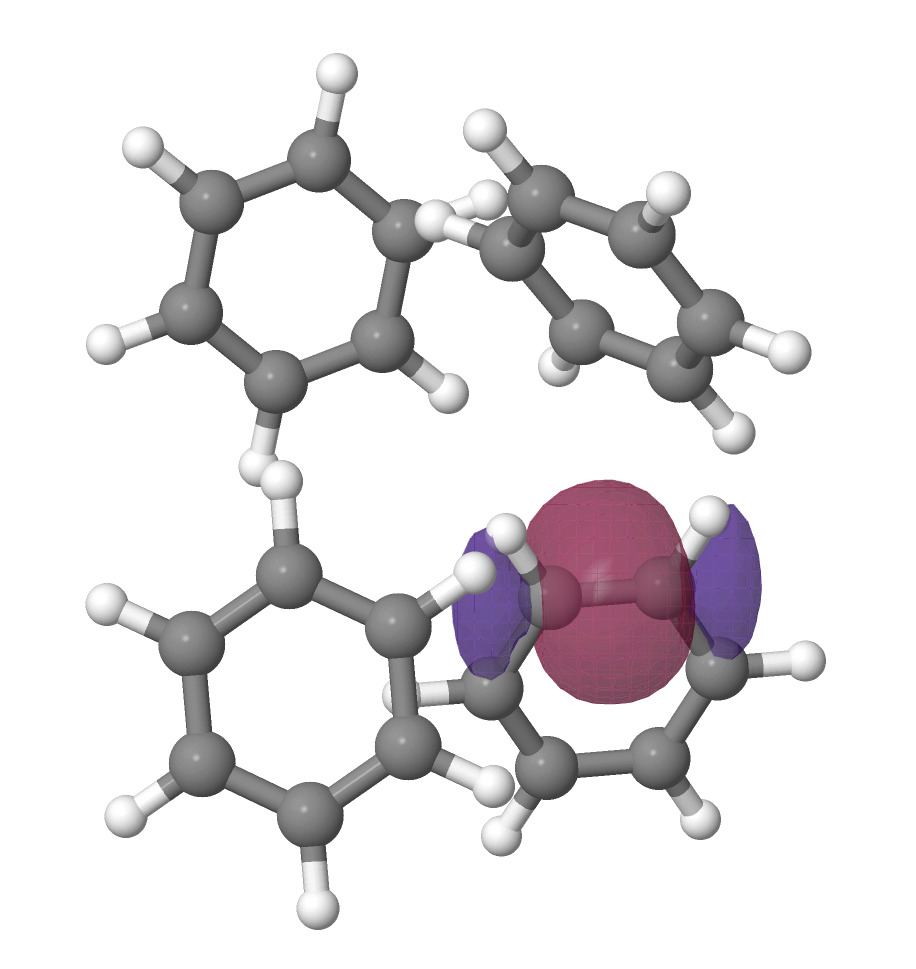}\includegraphics[width=0.33\textwidth]{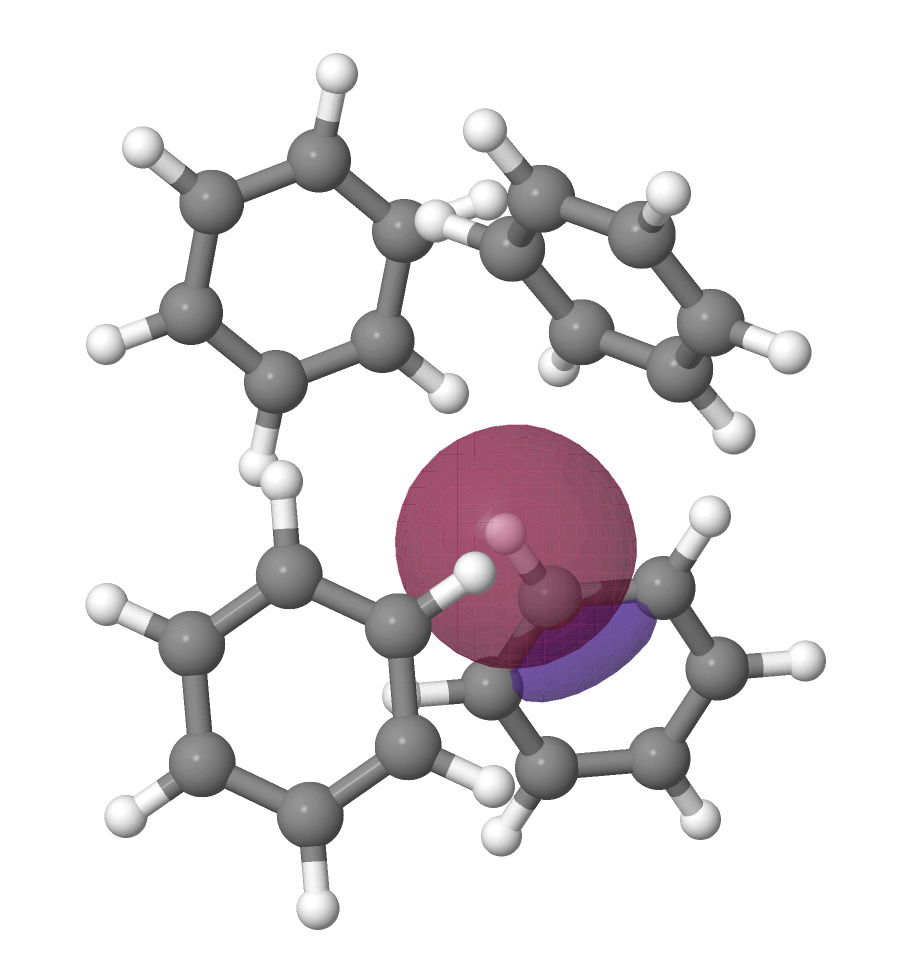}\includegraphics[width=0.33\textwidth]{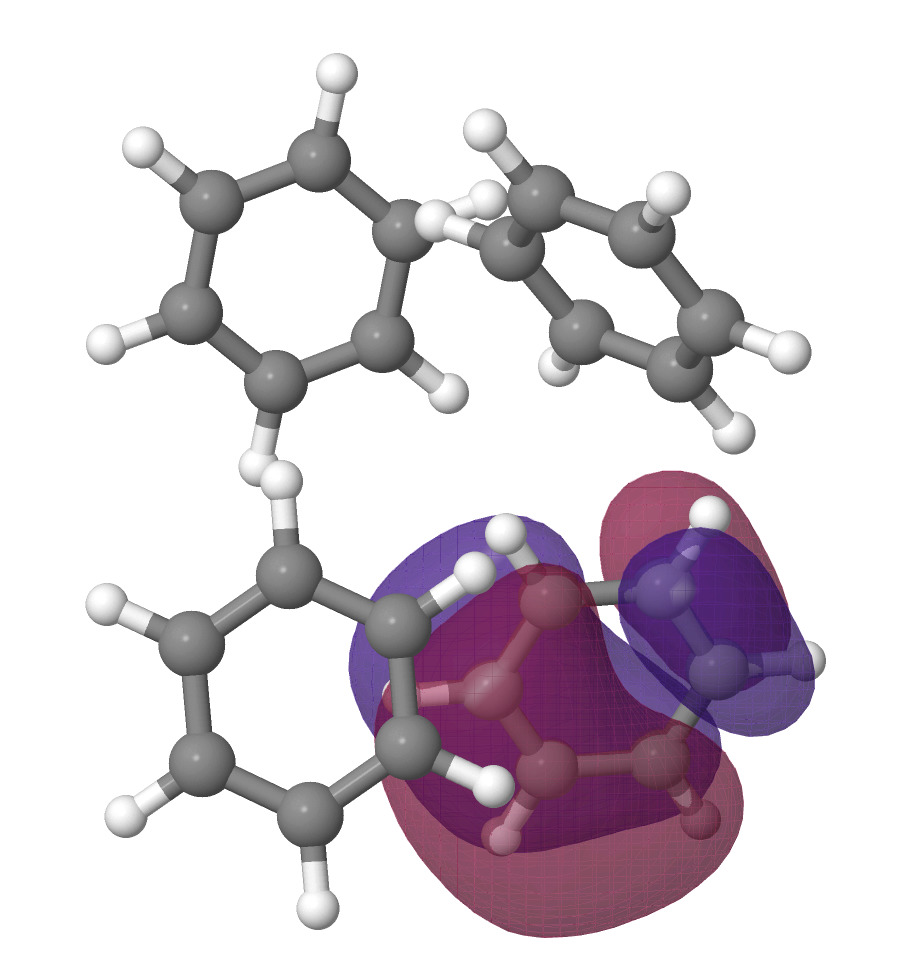} }

  \caption{Foster--Boys function (top row) and
    Pipek--Mezey Wannier function (bottom row) orbitals of the benzene
    crystal, represented by four benzene molecules in a
    three-dimensional periodic lattice (see main text). }
  \label{fig:benzene_crystal}
\end{figure*}

The assignment of $\sigma$- and $\pi$-bond character for the
COs is not as straightforward in the case of the benzene crystal
(\figref{benzene_crystal}), as there is no global plane of
symmetry. However, the resulting localized states associated with each
molecule in the crystal should conform to \eqref{sym-sigma, sym-pi},
unless they are a local mixture of $\sigma$ and $\pi$. Indeed,
a projection of the PMWFs about the molecular planes reveals that proper
$\sigma$-$\pi$ separation is maintained. Just as for an isolated
benzene molecule, each possible $\sigma_\mr{CC}$-, $\sigma_\mr{CH}$-
and $\pi_\mr{CC}$-bond in the crystal is described with highly
localized states. In contrast, the set of FBWFs consists once again of
twice the number of states with partial $\pi$ character, and half the
number of $\sigma_\mr{CC}$ states, compared to the COs, confirming that the
same mixing applies here as in the case of the other aromatic
hydrocarbons and cis-polyacetylene. Both methods produce
similar $\sigma_\mr{CH}$-bond orbitals (six in total), accounting
for all possible carbon-hydrogen bonds, in line with earlier
experience with the FB and PM methods\cite{Lehtola2013a, Lehtola2014}.

In all the cases discussed above, a clear qualitative difference exists
between the sets of PMWFs and FBWFs, with the latter method consistently
mixing $\sigma$- and $\pi$-bond orbitals in systems with alternating
single and double, or single and triple bonds.

For a periodic chain of polyethyl (-[C$_8$H$_{16}$]-),
and a sheet of graphene (not depicted), the resulting FBWFs and PMWFs
were found to be nearly identical in each case, although the FBWFs
were found to be more diffuse as judged by the isosurface for which
the integral of the orbital density\cite{Lehtola2014} reaches a value
of 0.75. Contrary to the aromatic hydrocarbons and cis-polyacetylene,
no mixing of $\sigma$ and $\pi$ states was found in the case of
graphene in the set of FBWFs.  A highly localized $\sigma$-bond exists
for each possible atom pair in both LO sets, and a localized
$\pi$-bond for every two carbon atoms in graphene.  Unlike in
graphene, the FBWFs in a sheet of boron nitride were found to consist
of a slight mix of $\sigma$- and $\pi$-bond COs.


\subsection{Localization measure of Pipek--Mezey orbitals}

To measure the localization of the orbitals obtained using the PMWF
method, we compare the values of the FBWF objective function
$\mathcal{L}$ (\eqref{FBWF-crit}) for COs, PMWFs, and FBWFs for all
systems studied here in \tabref{localization.FBWF}. While FBWFs
systematically yield the largest values of this objective function
(for which they have been optimized) -- indicating the largest degree
of locality -- the values obtained for the PMWFs are very close: the
difference
\begin{equation} \label{eq:dPMML}
  d = \frac{\mathcal{L}(\mathbf{W}^\text{PMWF}) - \mathcal{L}(\mathbf{W}^\text{FBWF})} {\mathcal{L}(\mathbf{W}^\text{FBWF})}
\end{equation}
is well under 1\% in every case. Relative to the localization measure
for the COs, this difference is even less significant, meaning that as
far as localization goes, the PMWF orbitals are practically as
spatially localized as the FBWF orbitals, as judged by the FBWF
criterion.

Similarly, the charge localization as measured by the PMWF objective
function $\mathcal{P}$ (\eqref{pbcpm}) is compared for COs, PMWFs, and
FBWFs, and the differences between PMWFs and FBMWs are calculated by
\eqref{dPMML} by reversing the roles of the orbitals, as now PMWFs are
by definition the most localized ones. The results are presented in
\tabref{localization.PMWF}. PMWFs yield the largest values, indicating
the largest degree of charge localization. However, the values
obtained using FBWFs are very close to the optimal values given by the
PMWFs, which states that FBWFs are practically as localized as PMWFs,
as judged by the PMWF criterion.  Thus, from the results in
\tabref{localization.FBWF, localization.PMWF} and the lack of a unique
definition for orbital locality we can conclude that PMWFs are just as
localized as FBWFs, which is one of the main results of this article.

In cases where $\sigma$-$\pi$ mixing is not an issue and the FBWF and
PMWF approaches give similar LOs, one might consider the FBWF method
superior to the PMWF approach due to its greater computational
efficiency: The evaluation of the FBWF cost function
(\eqref{FBWF-crit}) requires the formation of one $N \times N$ matrix,
$N$ being the number of occupied orbitals, whereas the evaluation of
the PMWF cost function (\eqref{pm}) requires the formation of $N_\mca$
atomic $N \times N$ partial charge matrices. However, in our
experience\cite{Lehtola2013a} the optimization of PM LOs converges
faster than that of FB LOs, and the larger cost of performing each
step in PM optimization compared to FB may be compensated by the fewer
number of iterations needed to optimize the PM cost function.  Indeed,
we have found this to be true also in the case of Wannier
functions.  \Tabref{numberofiterations} reports the number of
iterations $n_\text{iter}^\text{PMWF}$ and $n_\text{iter}^\text{FBWF}$
required to converge the PMWF and FBWF objective functions,
respectively. The numbers in \tabref{numberofiterations} have been
averaged over fifty separate optimization runs with randomly
initialized guesses for the localization matrix $\mathbf{W}$. The same
minimization method is used for both objective
functions\cite{Thygesen2005a}.  As can be seen from
\tabref{numberofiterations}, PMWF optimization typically requires up
to an order of magnitude fewer iterations to converge.

There are also ways in which PMWF could be made faster: for large
systems it's possible to reformulate the localization problem in terms
of some initial set of localized orbitals instead of the extended COs,
as this will make $\tilde{Q}^\mca_{RS}$ sparse. This initial set of
orbitals could be obtained e.g. via the Cholesky decomposition of the
density matrix\cite{Aquilante2006}. Approaches for the lossy
compression of $\tilde{Q}^\mca_{RS}$ could be pursued as well.
However, because the postprocessing to obtain LOs is typically an
insignificant portion of runtime compared to the actual KS-DFT
calculation to solve for the COs, we do not consider the potentially
larger computational effort for PMWF compared to FBWF to be an issue.

\begin{table}
 \centering
 \caption{Localization of the KS-DFT COs (KS), FBWFs and PMWFs as measured by the objective function value of \eqref{FBWF-crit}. } \small
  \begin{tabular*}{\columnwidth}{@{\extracolsep{\fill}} l | r | r | r | r }
    \hline
    System$^\dagger$ & $\mathcal{L}(\text{KS})$ & $\mathcal{L}(\text{FBWF})$ & $\mathcal{L}(\text{PMWF})$ & $d$ (\%) \\
    \hline
    b        & 12.86 & 14.42  & 14.40  & 0.10 \\
    co       & 37.61 & 52.38  & 52.31  & 0.13 \\
    sco      & 65.27 & 114.10 & 113.96 & 0.12 \\
    \hline
    cc $n=1$ & 6.85 & 15.43 & 15.39 & 0.22 \\
    cc $n=2$ & 6.26 & 31.58 & 31.56 & 0.07 \\
    cc $n=3$ & 7.56 & 47.70 & 47.68 & 0.02 \\
    cc $n=4$ & 9.22 & 63.77 & 63.76 & 0.01 \\
    \hline
    c-pa $n=1$ & 8.66 & 19.43 & 19.41 & 0.11 \\
    c-pa $n=2$ & 8.04 & 38.58 & 39.57 & 0.04 \\
    c-pa $n=3$ & 8.57 & 59.07 & 59.05 & 0.03 \\
    c-pa $n=4$ & 8.95 & 79.76 & 79.76 & 0.00 \\
    \hline
    ac(2,4) $n=1$ & 14.27 &  34.88 &  34.83 & 0.15 \\
    ac(2,4) $n=2$ & 13.48 &  71.12 &  71.08 & 0.06 \\
    ac(2,4) $n=3$ &  9.66 & 107.34 & 107.31 & 0.03 \\
    ac(2,4) $n=4$ &  8.81 & 143.49 & 143.46 & 0.02 \\
    \hline
    ac(3,3) $n=1$ & 19.58 &  50.40 &  50.30 & 0.20 \\
    ac(3,3) $n=2$ & 17.31 & 102.65 & 102.53 & 0.12 \\
    ac(3,3) $n=3$ & 13.88 & 154.95 & 154.85 & 0.07 \\
    ac(3,3) $n=4$ & 13.41 & 207.17 & 207.10 & 0.04 \\
    \hline
    ac(4,3) $n=1$ & 24.10 &  65.95 &  65.80 & 0.23 \\
    ac(4,3) $n=2$ & 22.95 & 134.12 & 133.93 & 0.14 \\
    ac(4,3) $n=3$ & 16.04 & 202.46 & 202.28 & 0.09 \\
    ac(4,3) $n=4$ & 12.73 & 265.28 & 264.73 & 0.21 \\
    \hline
    bn $n=(1,1)$ &  7.38 &  62.88 &  62.88 & 0.00 \\
    bn $n=(2,1)$ & 12.83 & 125.74 & 125.73 & 0.00 \\
    bn $n=(2,2)$ & 35.71 & 246.55 & 246.53 & 0.01 \\
    \hline
    gp $n=(1,1)$ & 7.39 &  62.08 &  62.03 & 0.08 \\
    gp $n=(2,1)$ & 5.63 & 126.06 & 125.97 & 0.07 \\
    gp $n=(2,2)$ & 1.84 & 249.73 & 249.53 & 0.05 \\
    \hline
    bc $n=(1,1,1)$ &  1.78 &  58.80 &  57.77 & 0.05 \\
    bc $n=(2,1,1)$ &  3.92 & 117.22 & 117.14 & 0.06 \\
    bc $n=(2,2,1)$ &  7.68 & 231.98 & 231.78 & 0.09 \\
    bc $n=(2,2,2)$ & 13.52 & 455.40 & 454.80 & 0.13 \\
    \hline
  \end{tabular*}\label{tab:localization.FBWF}
  $\dagger$ Abbreviations: benzene (b), coronene (co),
  supercoronene (sco), carbyne chain (cc), -[C$_8$]$_n$-; cis-polyacetylene chain (c-pa), -[C$_8$H$_8$]$_n$-; armchair nanoribbon chain (ac($i$,$j$) --
     width $i$ and length $j$) , [C$_{8i}$H$_8$]$_n$; boron nitride
     sheet (bn), [B$_{16}$N$_{16}$]$_{n=(x,y)}$; graphene sheet (gp),
     [C$_{32}$]$_{n=(x,y)}$; benzene crystal (bc),
     [C$_{24}$H$_{24}$]$_{n=(x,y,z)}$.
\end{table}

\begin{table}
 \centering
 \caption{Charge localization of the KS-DFT COs (KS), FBWFs and PMWFs
   as measured by the objective function value of \eqref{pbcpm}. } \small
 \begin{tabular*}{\columnwidth}{@{\extracolsep{\fill}} l | r | r | r | r }
   \hline
    System$^\dagger$ & $\mathcal{P}(\text{KS})$ & $\mathcal{P}(\text{FBWF})$ & $\mathcal{P}(\text{PMWF})$ & $d$ (\%) \\
    \hline
    b        & 2.01 &  6.10 &  6.16 & 0.82 \\
    co       & 2.64 & 21.24 & 21.46 & 1.01 \\
    sco      & 3.12 & 45.67 & 46.50 & 1.77 \\
    \hline
    cc $n=1$ & 1.64 &  6.66 &  6.70 & 0.59 \\
    cc $n=2$ & 1.57 & 13.44 & 13.54 & 0.71 \\
    cc $n=3$ & 1.55 & 20.22 & 20.37 & 0.75 \\
    cc $n=4$ & 1.55 & 27.02 & 27.19 & 0.61 \\
    \hline
    c-pa $n=1$ & 1.40 &  8.49 &  8.52 & 0.32 \\
    c-pa $n=2$ & 1.22 & 17.16 & 17.22 & 0.37 \\
    c-pa $n=3$ & 1.18 & 25.65 & 25.91 & 1.03 \\
    c-pa $n=4$ & 1.25 & 34.47 & 34.60 & 0.36 \\
    \hline
    ac(2,4) $n=1$ & 1.35 & 14.06 & 14.20 & 1.01 \\
    ac(2,4) $n=2$ & 1.11 & 28.38 & 28.67 & 1.03 \\
    ac(2,4) $n=3$ & 1.12 & 42.70 & 43.14 & 1.03 \\
    ac(2,4) $n=4$ & 1.05 & 57.11 & 57.60 & 0.84 \\
    \hline
    ac(3,3) $n=1$ & 1.49 & 19.82 & 20.19 & 1.88 \\
    ac(3,3) $n=2$ & 1.21 & 40.00 & 40.72 & 1.77 \\
    ac(3,3) $n=3$ & 1.11 & 60.24 & 61.26 & 1.67 \\
    ac(3,3) $n=4$ & 1.08 & 80.59 & 81.67 & 1.46 \\
    \hline
    ac(4,3) $n=1$ & 1.57 &  25.79 &  26.25 & 1.72 \\
    ac(4,3) $n=2$ & 1.24 &  51.90 &  52.86 & 1.82 \\
    ac(4,3) $n=3$ & 1.14 &  78.25 &  79.49 & 1.57 \\
    ac(4,3) $n=4$ & 1.09 & 103.78 & 105.24 & 1.39 \\
    \hline
    bn $n=(1,1)$ & 1.25 &  30.44 &  30.55 & 0.35 \\
    bn $n=(2,1)$ & 1.17 &  61.05 &  61.44 & 0.64 \\
    bn $n=(2,2)$ & 0.97 & 122.31 & 123.70 & 1.13 \\
    \hline
    gp $n=(1,1)$ & 0.87 &  24.07 &  24.11 & 0.19 \\
    gp $n=(2,1)$ & 0.81 &  48.34 &  48.50 & 0.32 \\
    gp $n=(2,2)$ & 0.68 & 101.21 & 101.75 & 0.54 \\
    \hline
    bc $n=(1,1,1)$ & 2.30 &  23.89 &  24.14 & 1.04 \\
    bc $n=(2,1,1)$ & 2.36 &  48.39 &  48.86 & 0.96 \\
    bc $n=(2,2,1)$ & 2.81 &  96.24 &  97.18 & 0.96 \\
    bc $n=(2,2,2)$ & 3.13 & 185.94 & 187.70 & 0.94\\
    \hline
    \hline
 \end{tabular*}\label{tab:localization.PMWF}
 $^\dagger$ See \tabref{localization.FBWF} for abbreviations.
\end{table}

\begin{table}
 \centering
   \caption{Number of iterations required to converge the PMWF
     objective function \eqref{pbcpm}, $n_\text{iter.}^\text{PMWF}$,
     and the FBWF objective function \eqref{FBWF-crit},
     $n_\text{iter.}^\text{FBWF}$. The number of iterations are
     averaged over fifty independent runs starting with randomly
     initialized guesses for the rotation matrices $\mathbf{W}$.} \small
  \begin{tabular*}{\columnwidth}{@{\extracolsep{\fill}} l | r | r }
    \hline
    System$^\dagger$ & $n_\text{iter.}^\text{PMWF}$ & $n_\text{iter.}^\text{FBWF}$ \\
    \hline
    b        & 49 & 83  \\
    co       & 81 & 107 \\
    sco      & 94 & 295 \\
    \hline
    cc $n=1$ & 686 & 1398 \\
    cc $n=2$ & 799 & 3509 \\
    cc $n=3$ & 786 & 4321 \\
    cc $n=4$ & 733 & 4177 \\
    \hline
    c-pa $n=1$ & 722 &  682 \\
    c-pa $n=2$ & 885 & 1659 \\
    c-pa $n=3$ & 937 & 2753 \\
    c-pa $n=4$ & 925 & 2609 \\
    \hline
    ac(2,4) $n=1$ & 169 &  817 \\
    ac(2,4) $n=2$ & 253 & 2283 \\
    ac(2,4) $n=3$ & 285 & 3475 \\
    ac(2,4) $n=4$ & 314 & 4177 \\
    \hline
    ac(3,3) $n=1$ & 101 & 1069 \\
    ac(3,3) $n=2$ & 180 & 2913 \\
    ac(3,3) $n=3$ & 272 & 4377 \\
    ac(3,3) $n=4$ & 355 & 4511 \\
    \hline
    ac(4,3) $n=1$ & 125 & 1160 \\
    ac(4,3) $n=2$ & 184 & 2690 \\
    ac(4,3) $n=3$ & 216 & 4269 \\
    ac(4,3) $n=4$ & 298 & 4672 \\
    \hline
    bn $n=(1,1)$ & 183 & 2067 \\
    bn $n=(2,1)$ & 223 & 3322 \\
    bn $n=(2,2)$ & 281 & 4183 \\
    \hline
    gp $n=(1,1)$ & 123 & 1208 \\
    gp $n=(2,1)$ & 179 & 2702 \\
    gp $n=(2,2)$ & 214 & 3411 \\
    \hline
    bc $n=(1,1,1)$ & 610 & 1491 \\
    bc $n=(2,1,1)$ & 754 & 3117 \\
    bc $n=(2,2,1)$ & 819 & 3892 \\
    bc $n=(2,2,2)$ & 769 & 4018 \\
    \hline
  \end{tabular*}\label{tab:numberofiterations}
  $^\dagger$ See \tabref{localization.FBWF} for abbreviations.
\end{table}


\section{Summary and Discussion}

An extension of the generalized Pipek--Mezey method of
\citeref{Lehtola2014} for the formation of Wannier functions with the
projector augmented wave formalism has been presented, as well as an
implementation that supports $\bk$-point sampling and multiple
possible representations for the electronic wave function: plane
waves, real-space grids, and linear combination of atomic orbitals.
Applications of the method to a variety of different systems have also
been presented, ranging from isolated molecules to periodic systems in
one, two, and three dimensions. The Pipek--Mezey Wannier functions
(PMWFs) have been compared to the commonly used ``maximally localized
Wannier functions''. However, since there is no unique, unambiguous
way of defining the locality of orbitals and several possible measures
for orbital locality have been defined in the literature (the orbitals
corresponding to a given objective function are maximally localized
\emph{as determined by that objective function}), we choose to refer
to Foster--Boys Wannier functions (FBWFs) instead of ``maximally
localized Wannier functions''.

PMWFs are just as highly localized as the FBWFs, as revealed by
cross-comparison of the values of the objective functions of one of
the measures evaluated with the orbitals from the other.  The PMWFs
do, however, offer an advantage over FBWFs in that a clear separation
is obtained between $\sigma$- and $\pi$-bond orbitals. In the majority
of the systems studied here, some of the FBWFs turned out to be a mix
of $\sigma$ and $\pi$ orbitals resulting in a less clear chemical
interpretation. The PMWF method gives localized orbitals that are
consistent with chemical intuition in the number of single, double,
and triple bonds for carbon and hydrocarbon systems.

Recently, Pipek--Mezey orbitals have been used in molecular
calculations, e.g., in studying hydrogen transfer in aryloxy
radicals\cite{Chen2010}, ring currents in aromatic
molecules\cite{Gibson2014}, hydrogenolysis of nickel-methyl
bonds\cite{Curado2014}, bonding in an amino-borane rhodium
complex\cite{Kumar2016}, as well as electron flow in reaction
mechanisms\cite{Knizia2015}. The introduction of the PMWF method for
condensed matter simulations using periodic boundary conditions opens
up new possibilities for the chemical interpretation of, e.g., surface
chemistry, where the mixing of $\sigma$ and $\pi$ states in the FBWF
approach may cause problems for the interpretation of which orbitals
of the surface are participating in the reaction, as $\sigma$ and
$\pi$ electrons typically react in a different way\cite{Knizia2015}. Thus, we
expect the procedure outline here to become widely used in a variety
of applications, including theoretical studies of heterogeneous catalysis.

In addition to their use for visualization and interpretation
purposes, PMWFs can also be useful for local post-HF
methods\cite{Pulay1983, *Saebo1993}.  While a FBWF implementation of
local M\o{}ller--Plesset perturbation theory\cite{Moller1934}
truncated at the second order (MP2) has been
reported\cite{Casassa2006, *Pisani2008}, an implementation based on
PMWFs can be expected to have better performance, just as the PM
approach has been found to be superior to FB for local treatment of
electron correlation effects in molecules\cite{Boughton1993}.

%

\section*{Acknowledgments}
Computational resources provided by CSC -- IT Center for Science,
Ltd. (Espoo, Finland) are gratefully acknowledged.  This work was
funded by the Academy of Finland trough its Centres of Excellence
Programme (2012--2017) under Project No. 251748, and through its
FiDiPro Programme under Project No. 263294.

\section{Appendix}
The extension of the generalized overlap matrices under periodic
boundary conditions from the $\Gamma$-point to ${\bf k}$-point
sampling is described in this appendix.  Given a unit cell with basis
vectors ${\bf a}_1$, ${\bf a}_2$ and ${\bf a}_3$, the reciprocal
lattice vectors ${\bf b}_1$, ${\bf b}_2$ and ${\bf b}_3$ are
defined. Assuming a uniform sampling of the first Brillouin zone, any
${\bf k}$-point can be expressed as
\begin{equation}
 {\bf k} = \frac{n_1}{N_1}{\bf b}_1 + \frac{n_2}{N_2}{\bf b}_2
           + \frac{n_3}{N_3}{\bf b}_3
\end{equation}
where $N_i$ is the number of ${\bf k}$-points in the direction of
${\bf b}_i$, and $n_i=0,\dotsc,N_i-1$. The Bloch states $\psi_{n{\bf k}}$
correspond to the $\Gamma$-point eigenstates of the repeated cell
defined by the extended basis vectors $N_i{\bf a}_i$. The reciprocal lattice
vectors, ${\bf G}_\alpha$ and corresponding weights, $g_\alpha$,
in \eqref{pbcpm, pbccopcm},
now refer to the extended basis vector defined by
${N_1}{\bf a}_1$, ${N_2}{\bf a}_2$,${N_3}{\bf a}_3$.

The weights
associated with the objective function values of \eqref{pbcpm} and
\eqref{FBWF-crit} are normalized such that
\begin{equation}
 g_\alpha = \frac{g'_\alpha}{\sum_\alpha^{N_\alpha}g'_\alpha}
\end{equation}
With this definition of weights the objective function values of \eqref{pm}
and \eqref{pbcpm} are in agreement between calculations employing open boundaries
and those with periodic boundaries (say for a molecule in vacuum).
Furthermore, the PMWF and FBWF objective function values, \eqref{pbcpm} and \eqref{FBWF-crit}, respectively, also become systematic with system size.

In the PAW approach\cite{Blochl1994} the all-electron (AE) wave
functions (WFs) $\psi_s$ are represented in terms of smooth pseudo
(PS) waves $\ti{\psi}_s({\bf r})$
\begin{equation}
  \psi_s({\bf r}) = \ti{\mathcal{T}}\ti{\psi}_s({\bf r}).
\end{equation}
The transformation operator $\ti{\mathcal{T}}$ is given by
\begin{equation}
  \ti{\mathcal{T}} = 1 + \sum_\mca \sum_i \left(\phi^\mca_i({\bf r}) -
  \ti{\phi}^\mca_i({\bf r})\right) \bra{\ti{p}^\mca_i}
\end{equation}
where $\phi^\mca_i$ and $\ti{\phi}^\mca_i$ are AE and PS (smooth)
partial waves. The partial waves are equal outside atom-centered
augmentation spheres of radii $r^\mca_c$
\begin{equation}
  \phi^\mca_i({\bf r}) = \ti{\phi}^\mca_i({\bf r}),\quad \left|{\bf r}
  - {\bf R}^\mca \right| > r_c^\mca
\end{equation}
where ${\bf R}^\mca$ is the position of atom $\mca$. $\ti{p}^\mca_i$
are projector functions, which determine the expansion coefficients of
the PS WFs inside the augmentation region. The projector functions are
atom centered
\begin{equation} \label{eq:PAW-proj}
  \ti{p}^\mca_i({\bf r}) = \ti{p}^\mca_{n_i l_i}(\left|{\bf r} - {\bf
    R}^\mca\right|)Y_{l_i m_i}\left( \frac {{\bf r} - {\bf R}^\mca} {\left|{\bf r} - {\bf R}^\mca \right|} \right),
\end{equation}
where $Y_{lm}$ are spherical harmonics.

The AE WFs are given by
\begin{equation} \label{eq:PAW-AE}
  \psi_s({\bf r}) = \ti{\psi}_s({\bf r}) +
  \sum_\mca \sum_i \left(\phi^\mca_i({\bf r}) - \ti{\phi}^\mca_i({\bf
    r})\right) \braket{\ti{p}^\mca_i | \ti{\psi}_s},
\end{equation}
where the sum over $i$ indicates a sum over the principal quantum
numbers $n,l,m$ in the PAW expansion as given in
\eqref{PAW-proj}. However, the representation of \eqref{PAW-AE} is
rarely used since it requires an extremely fine grid to properly
describe the AE partial wave due to rapid oscillations of
$\phi^\mca_i({\bf r})$ close to ${\bf R}^\mca$.

Instead, within the PAW approach, the expectation value of an operator
$A$, $\braket{A} = \sum_s f_s \braket{\psi_s | A | \psi_s}$ is
expressed in terms of the PS states and a corresponding PS operator
$\ti{A}$: $\braket{A} = \sum_s f_s \braket{\ti{\psi}_s | \ti{A} |
  \ti{\psi}_s}$. A quasilocal operator, such as the real-space
projection operator $\ket{\bf r} \bra{\bf r}$, acting on the PS states
has the following form
\begin{align}\label{eq:psoperator}
  \ti{A} =& \ti{\mathcal{T}}^\dagger A\ti{\mathcal{T}} \nonumber \\ =&
  A + \sum_\mca \sum_{i_1i_2} \ket{\ti{p}_{i_1}^\mca} \left(
  \braket{\phi_{i_1} | A | \phi_{i_2}} - \braket{\ti{\phi}_{i_1}^\mca | A
    | \ti{\phi}_{i_2}^\mca} \right) \bra{\ti{p}_{i_2}^\mca}.
\end{align}
The overlap operator $S_{ij} = \braket{i | j}$ is
\begin{equation} \label{eq:pawS}
  \hat{S} = \ti{\mathcal{T}}^\dagger\ti{\mathcal{T}} = 1 + \sum_\mca
  \sum_{i_1i_2} \ket{\ti{p}_{i_1}^\mca} \Delta S^\mca_{i_1i_2}
  \bra{\ti{p}_{i_2}^\mca}
\end{equation}
where
\begin{equation} \label{eq:deltaS}
  \Delta S_{i_1i_2}^\mca = \braket{\phi_{i_1}^\mca | \phi_{i_2}^\mca}
  - \braket{\ti{\phi}_{i_1}^\mca | \ti{\phi}_{i_2}^\mca}.
\end{equation}
The PS WFs are orthonormal
\begin{equation}\label{eq:overlap}
  \braket{\ti{\psi}_r | \hat{S} | \ti{\psi}_s} = \delta_{rs}
\end{equation}
only with respect to the PAW overlap operator given in
\eqref{pawS}. This forms the basis for the following overlap schemes
to define the partial charge matrices $Q_{mn}^\mca$ that are used for
the generalized Pipek--Mezey localization method.

Using \eqref{psoperator, overlap} and the assumption that the atomic
weight function centered on $\mca$ has negligible weight in the
augmentation region around atom $\mca'$, i.e.
\begin{equation}\label{eq:condition2}
  w_\mca({\bf r})| {\bf r} \in \mca' = \begin{cases}
    1 & \quad \mr{if\, \,} \mca = \mca' \\
    0 & \quad \mr{if\, \,} \mca \neq \mca'
  \end{cases}
\end{equation}
the PS partial charge projection operator, $\hat{S}_w^\mca$, becomes
\begin{align}
  \hat{S}_w^\mca =& \ti{\mathcal{T}}w_\mca\ti{\mathcal{T}} \nonumber \\
  =& w_\mca + \sum_{\mca' \mca''} \sum_{i_1i_2} \ket{\ti{p}_{i_1}^{\mca'}} \bra{\ti{p}_{i_2}^{\mca''}} \nonumber \\
  \times & \left( \braket{\phi_{i_1}^{\mca'} | w_\mca | \phi_{i_2}^{\mca''}}
  -  \braket{\ti{\phi}_{i_1}^{\mca'}  | w_\mca | \ti{\phi}_{i_2}^{\mca''}}
  \right) \nonumber  \\
  =& w_\mca + \sum_{i_1i_2} \ket{\ti{p}_{i_1}^\mca} \Delta S^\mca_{i_1i_2}
  \bra{\ti{p}_{i_2}^\mca}
\end{align}
where in the last line the condition of \eqref{condition2} is used,
and the resulting partial charge matrix is simply
\begin{equation}
  Q_{rs}^\mca = \braket{ \ti{\psi}_r | w_\mca | \ti{\psi}_s} +
  \sum_{i_1i_2} \braket{ \ti{\psi}_r | \ti{p}_{i_1}^\mca} \Delta
  S^\mca_{i_1i_2} \braket{\ti{p}^\mca_{i_2} | \ti{\psi}_s}.
\end{equation}
The first term on the right hand side is given by
\begin{equation}
  \braket{ \ti{\psi}_r | w_\mca | \ti{\psi}_s } = \int
  \ti{\psi}^*_r({\bf r})w_\mca({\bf r})\ti{\psi}_s({\bf r}) {\rm d}^3r.
\end{equation}

In the case of periodic systems the partial charge matrix in the CO
basis is
\begin{align}
  Q_{\alpha,RS}^\mca = & \braket{ \ti{\psi}_R | e^{-{\bf G}_\alpha\cdot{\bf r}} w_\mca | \ti{\psi}_S}  + \sum_{i_1 i_2} \braket{\ti{\psi}_R | \ti{p}_{i_1}^\mca}   \braket{\ti{p}^\mca_{i_2} | \ti{\psi}_S} \nonumber \\
  \times & \left( \braket{ \phi_{i_1}^\mca | e^{-{\bf G}_\alpha\cdot{\bf r}} | \phi_{i_2}^\mca} - \braket{\ti{\phi}_{i_1}^\mca | e^{-{\bf G}_\alpha\cdot{\bf r}} | \ti{\phi}_{i_2}^\mca} \right)
\end{align}
Now, assuming the phase of the exponential does not vary significantly
over the space where $\ti{p}^\mca_{i}$ is nonzero (this approximation
is also used in the Foster--Boys Wannier function PAW
analog\cite{Thygesen2005a}, and more generally, with ultra-soft
pseudopotentials\cite{Bernasconi2001}), the integral in the last term
can be estimated by
\begin{equation}
  e^{-{\bf G}_\alpha\cdot{\bf R}^\mca}\sum_{i_1i_2} \braket{\ti{\psi}_R |
    \ti{p}_{i_1}^\mca} \Delta S^\mca_{i_1i_2} \braket{\ti{p}^\mca_{i_2} |
    \ti{\psi}_S},
\end{equation}
where we have used the locality of the atomic PAW projectors
(\eqref{condition2}). We also note that the weight functions in a cell
$n$ satisfy
\begin{equation}
  w_{\mca,n}({\bf r} - {\bf R_n}) = w_{\mca,0}({\bf r})
\end{equation}
where 0 indicates the cell at the origin, and ${\bf R}_n$ is a Bravais
lattice vector.

The numerical stability of the Hirshfeld-type partitioning function,
\eqref{hirsh}, using the Gaussian model densities is further enforced
by employing a cut-off radius, $R_c$, such that
\begin{equation}\label{eq:gausscoff}
  \overline{n}^c_\mca(\br) = \begin{cases} \overline{n}_\mca(\br) &
    \quad \text{if $\left| \br - {\bf R}^\mca \right| \leq R_c$} \\ 0 &
    \quad \text{otherwise}
  \end{cases}
\end{equation}
and we use $\overline{n}^c_\mca(\br)$ to evaluate \eqref{hirsh}. A
constant cut-off of 3.8 \AA\ is applied for all types of atoms.
Although the partial atomic charge will change when applying a
different cut-off, this parameter is not explored further, based on
the experience from \citeref{Lehtola2014} and the fact that the
different choices for the decay parameters result in similar LOs.

Finally, as the real-space grid used to represent the wave functions
(coarse grid in GPAW) is uniform, it can lead to problems when the
atoms of the system are also distributed uniformly when the
Wigner--Seitz (WS) weight function is constructed. For instance,
assume two atoms in the system to be placed on the $z$ axis, and the
wave function grid spacing to be $\Delta$. Now, if the interatomic
separation $R$ is an even multiple of $\Delta$: $\exists n \in
\mathcal{N}: R=2n\Delta$, then the grid points on the $z=n\Delta$
plane will be at the same distance from both nuclei, which would lead
the points to contribute with unit weight to both atomic regions,
breaking the condition in \eqref{w-crit2}. Hence, the WS weight
function is modified to rectify this problem by redistributing the
weight evenly to all the atoms that share the point as
\begin{equation}\label{eq:WS-normalized}
  w_\mca^\mr{WS}({\bf r}) = \left[   \sum_{\mca': {\bf r} \in \mca'} 1 \right]^{-1},
\end{equation}
after which the sum of weights will be unity, once again satisfying
\eqref{w-crit2}.

\clearpage{}


\end{document}